\documentclass[journal]{IEEEtran}
\usepackage{amsfonts,amssymb,amsmath,amsbsy,bm,paralist,theorem,ifthen,color}
\usepackage{graphics,graphicx,subfigure}
\usepackage{epstopdf}
\usepackage{fancyhdr}
\usepackage{makecell}
\usepackage{url}
\usepackage{multirow}

\ifCLASSINFOpdf
\else
\fi
\begin{document}
%
\title{Interference Cancellation Based Channel Estimation for Massive MIMO Systems with Time Shifted Pilots}
%
%
%

\author{\;Bule Sun,
        \;Yiqing Zhou,~\IEEEmembership{Senior Member,~IEEE,}
        \;Jinhong Yuan,~\IEEEmembership{Fellow,~IEEE,}
        \;Jinglin Shi,~\IEEEmembership{Member,~IEEE}
\thanks{\textcircled{c}2020 IEEE. Personal use of this material is permitted. Permission from IEEE must be obtained for all other uses, in any current or future media, including reprinting/republishing this material for advertising or promotional purposes, creating new collective works, for resale or redistribution to servers or lists, or reuse of any copyrighted component of this work in other works.}
\thanks{This work was supported by the National Natural Science Foundation of China under Grant 61571425.}
\thanks{B. Sun, Y. Zhou, J. Yuan and J. Shi are with Beijing Key Laboratory of Mobile Computing and Pervasive Device and the Institute of Computing Technology, Chinese Academy of Sciences, Beijing 100190, China, and also with the University of Chinese Academy of Sciences, Beijing 100049, China; J. Yuan is also with the School of Electrical Engineering and Telecommunications, University of New South Wales, Sydney, NSW, 2052 Australia. (e-mail: sunbule@ict.ac.cn; zhouyiqing@ict.ac.cn; j.yuan@unsw.edu.au; sjl@ict.ac.cn.)}
\thanks{\it{(Corresponding author: Yiqing Zhou.)}}
\thanks{Manuscript received 30-Jul-2019; revised 04-Jan-2020, 24-Apr-2020,  and 20-Jun-2020; accepted 21-Jun-2020.}
}

%
%

\markboth{IEEE Transactions on Wireless Communications,~Vol.~xx, No.~xx, 2020}%
{Shell \MakeLowercase{\textit{et al.}}: Interference Cancellation Based Channel Estimation for Massive MIMO Systems with Time Shifted Pilots}

%



\maketitle

\begin{abstract}
In massive multiple-input multiple-output (MIMO) systems with time shifted pilot (TSP) schemes, the inter-group interference caused by the pilot contamination can be eliminated when the number of base station (BS) antennas $M$ approaches infinity. However, $M$ is finite in practice and the effectiveness of the TSP is limited by channel estimation errors. In this paper, it is analytically shown that the mean square channel estimation error (MSCEE) of the TSP is dominated by the inter-group data interference. To reduce the MSCEE in the finite antenna massive MIMO systems, an interference cancellation based channel estimation for the TSP (IC-TSP) is proposed, where the dominant inter-group data interference is canceled based on BS cooperation. To show the advantage of the IC-TSP, the additional overhead of IC-TSP is evaluated by considering different $M$ and the coherence time of BS-BS channels. Furthermore, the impact of sectorization and compressed sensing based BS-BS channel estimation are also discussed. We show that when $128\le M\le 2048$, with the inter-group data interference from the nearest two cell layers being canceled, the IC-TSP achieves a spectral efficiency gain of more than 1.2 bps/Hz over the TSP.
\end{abstract}


%
\IEEEpeerreviewmaketitle

\section{Introduction}
%
%
%
%
\IEEEPARstart{M}{assive} multiple-input multiple-output (MIMO) is a promising candidate for the fifth generation (5G) or beyond 5G mobile communication system [1]-[7]. The main idea of massive MIMO is to deploy a large number of antennas at base stations (BSs), i.e., $M$, to serve a small number of mobile stations (MSs), i.e., $K$ ($M\gg K$). Under favorable propagation conditions, simple linear precoding and detecting methods are able to achieve significant gains in throughput compared with conventional MIMO systems, where channel estimation is needed. Due to the large number of antennas at the BS, the amount of pilots needed in downlink (DL) channel estimation is huge. In contrast, resources needed for uplink (UL) channel estimation are much less since the number of MSs is relatively small. Exploiting the channel reciprocity of time division duplex (TDD) transmission mode, the information of DL channel can be obtained from UL channel estimation, which is not easy in frequency division duplex (FDD) systems. However, even with TDD, massive MIMO faces serious pilot contamination [1]. This occurs because the time-frequency resource to carry pilots for channel estimation is limited, and different cells have to reuse the same resource which results in serious inter-cell interference (ICI) [1].

A number of studies have been carried out to tackle the pilot contamination problem. One straightforward solution is to avoid using pilot for channel estimation, i.e., the blind channel estimation [8]. However, it is difficult to be deployed in practice since the complexity increases proportionally to ${{M}^{2}}$. For pilot-based channel estimation, there are two pilot contamination reduction approaches, i.e., aligned pilot (AP) based and time shifted pilot (TSP) based methods [9]. For AP based methods, MSs in different cells transmit UL pilots using the same time-frequency resource. Various schemes have been proposed to mitigate the pilot contamination for the AP based methods [10]-[13]. However, due to the synchronized receptions/transmissions among different cells at both pilot and data transmission stage, the AP scheme actually stands for the worst case of TSP in terms of spectral efficiency [1]. This is because the ICI during data transmission is highly correlated with the channel estimation error caused by pilot contamination. The ICI will be significantly aggravated when using precoding or detection based on this polluted channel estimation. The TSP is proposed in [9], separating the transmission of pilot signals in different cells on different time resources of one coherence time. Due to the limited length of coherence time, the same time resources must be reused for pilot in different cells, similar to the frequency reuse. Define a cell cluster composed of adjacent cells with orthogonal resources for pilot, and a cell group including all the cells using the same resources for pilot transmission. With TSP, MSs in one cell group transmit UL pilots while other cell groups are transmitting DL data. Therefore, the UL pilot in one cell is contaminated by the UL pilot from the same cell group (i.e., intra-group interference) and DL data from all other cell groups (i.e., inter-group interference). Based on the channel estimated at UL, precoding can be carried out at the BS to achieve good performance in DL transmission. It has been demonstrated in [9] that in a massive MIMO system with infinite number of BS antennas, the inter-group interference can be smartly canceled out by exploiting the asymptotic channel orthogonality.

Note that, current massive MIMO testbeds and commercial products can only support no more than 256 antennas due to the limitation of hardware [14]-[19]. It is expected that in practice, massive MIMO systems can only employ limited number of antennas, e.g., less than 10,000 for quite a long time. For a practical massive MIMO system, the previously discussed inter-group interference is not negligible [20] and it increases significantly with the channel estimation error. To reduce the channel estimation error, a receive beamforming (RBF) method based on the orthogonal basis decomposition is proposed in [21], where the RBF projects the pilot signal to the orthogonal space of the UL data, eliminating the interference from UL data transmission. However, using TSP, the pilot is mainly interfered by DL data transmission in nearby cells, but not UL data transmission. So the performance improvement of [21] is limited. Therefore, considering TSP with finite antennas, it is important to develop effective methods to improve the performance of channel estimation.

Considering a TDD massive MIMO system with TSP, this paper targets to improve the channel estimation accuracy for massive MIMO systems with a finite number of BS antennas $M$. The main contributions of our work are summarized as follows.

\begin{itemize}
    \item The mean square channel estimation error (MSCEE) is analyzed with finite $M$. We show that the MSCEE of the TSP is determined by the inter-group data interference, i.e., the ICI from DL data transmission in other groups.
    \item We derive the DL and UL signal to interference plus noise ratio (SINR) for the TSP massive MIMO with finite $M$. We prove that the impact of the MSCEE on the SINR is significant when $M$ is finite. To achieve a practical target SINR $SIN{{R}_{\Upsilon }}$, the number of BS antenna needed for the TSP, ${{M}_{T}}$, is analytically described. In particular, we show that ${{M}_{T}}$ increases rapidly with the MSCEE with a steep slope, which is inversely proportional to the large scale fading of target MS.
    \item We propose an interference cancellation (IC) based channel estimation for TSP (IC-TSP) to reduce the MSCEE. The basic idea is to cancel the dominant inter-group DL data interference by using BS cooperation. We demonstrate that the proposed IC-TSP can reduce the MSCEE by 15 dB (with proper system settings) and achieves a spectral efficiency gain of more than 1.2 bps/Hz over TSP when $128\le M\le 2048$.
    \item For IC-TSP, we evaluate the impact of the additional pilot overhead on the spectral efficiency by considering different coherence time of BS-BS channels and BS antenna number $M$. To achieve higher effective SINRs than the TSP, the IC-TSP needs a BS-BS channel coherence time longer than a specific value, to compensate the overhead introduced by BS-BS channel estimation. Since both the SINR and the pilot overhead increases as $M$ increases, there exist an optimal value for $M$ maximizing the spectral efficiency for the IC-TSP. Furthermore, when $M$ is sufficiently large, it is possible that spectral efficiency of IC-TSP become lower than that of TSP. We also evaluate the impact of sectorization and the compressed sensing (CS) based BS-BS channel estimation on the spectral efficiency of IC-TSP. Both these two approaches are more beneficial when $M$ is large due to the significantly reduced pilot overhead.
\end{itemize}

Note that the initial idea of our proposed methodology is presented in [22]. Different to [22], this paper analyzes the dominant component of MSCEE and studies the impact of the MSCEE on the SINR of TSP, which demonstrates the importance to improve the channel estimation quality. Furthermore, the advantage of the IC-TSP is strengthened by combining the IC-TSP with the sectorization and the CS based BS-BS channel estimation. Overall, this paper presents a further comprehensive study based on our initial research in [22].

The rest of the paper is organized as follows. In Sec. II, the system model is described. In Sec. III, with finite BS antennas, the MSCEE in TSP is derived and its impacts on DL and UL SINR are evaluated. Then the IC-TSP is proposed in Sec. IV, where the impact of system parameters and pilot overhead reducing approaches are also analyzed. Simulation results are presented in Sec. V. Finally, conclusions are drawn in the last section.

Throughout the paper, $\mathbf{A}\in {{\mathbb{C}}^{M\times N}}$ denotes an $M\times N$ complex matrix. ${{\left( \mathbf{A} \right)}^{*}}$, ${{\left( \mathbf{A} \right)}^{T}}$ and ${{\left( \mathbf{A} \right)}^{H}}$ represent the conjugate, transpose and conjugate transpose of matrix $\mathbf{A}$, respectively. $\left\| \mathbf{a} \right\|$ denotes the Euclidean norm of vector $\mathbf{a}$, ${{\mathbf{I}}_{N}}$ is the $N\times N$ identity matrix, and ${{\mathbf{0}}_{N}}$ denotes all-zero $N\times 1$ vector. ${\bf{n}} \sim {\cal C}{\cal N}\left( {{\bf{a}},{\rm{ }}{\bf{A}}} \right){\rm{ }}$ is a complex Gaussian vector with mean $\mathbf{a}$ and covariance matrix $\mathbf{A}$. $\mathbb{E}\left\{ \cdot  \right\}$ and $\mathbb{D}\left\{ \cdot  \right\}$ denote the operation to get expectations and variances, respectively. $\Re \left( \cdot  \right)$ and $\Im \left( \cdot  \right)$ denote the operation to get the real and imaginary parts, respectively. $[\mathbf{A}]_{p,q}$  denotes the $\left( p,q \right)$-th element of matrix $\mathbf{A}$. $|S|$ denotes the number of elements in set $S$.

\section{System Model}
\begin{figure*}[htbp]
  \centering
  \includegraphics[angle=0,width=0.95\textwidth]{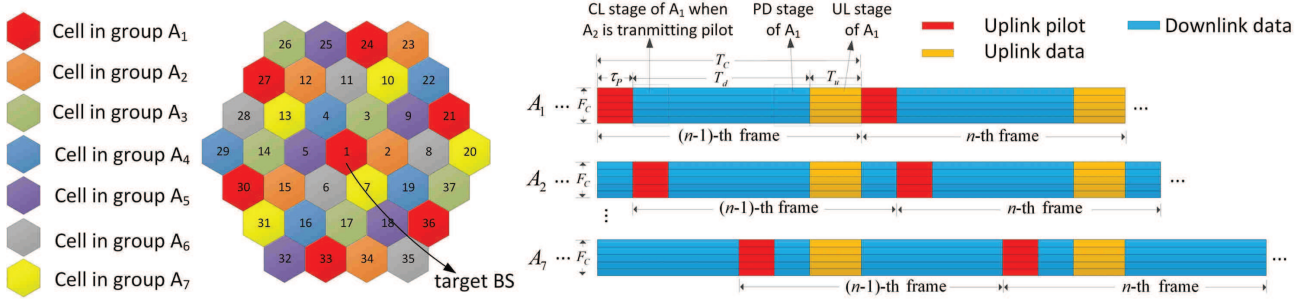}\\
  \caption{Illustration of cell grouping and the TSP transmission with $\Gamma=7$ and $L=37$.}\label{Fig. 1}
\end{figure*}

\newcounter{TempEqCnt}
\setcounter{TempEqCnt}{\value{equation}}
\setcounter{equation}{0}
\begin{figure*}[ht]
\begin{equation}
{{\mathbf{y}}_l} \!=\! \underbrace {\sum\limits_{k = 1}^K {\sqrt {\rho _{UL,lk}^P} {{\mathbf{g}}_{llk}}{{\mathbf{\psi }}_k}} }_{{\text{pilot from target cell}}}\!{+}\!\!\underbrace {\sum\limits_{j \ne l,j \in {A_p}}^{} {\sum\limits_{k = 1}^K {\sqrt {\rho _{UL,jk}^P} {{\mathbf{g}}_{ljk}}{{\mathbf{\psi }}_k}} } }_{{\text{ intra-group pilot interference}}}\! +\!\! \underbrace {\sum\limits_{d = 1,d \notin {A_p}}^L \!\!\!\!\!{{{\mathbf{G}}_{ld}}\sum\limits_{k = 1}^K {\sqrt {\rho _{DL,dk}^D} {{\mathbf{w}}_{dk}}{\mathbf{x}}_{dk}^D} } }_{{\text{inter-group DL data interference}}} \!+ \!\underbrace {{{\mathbf{n}}_{P,l}}}_{{\text{noise}}},
\end{equation}
\hrulefill
\vspace{-0.5cm}
\end{figure*}
\setcounter{equation}{\value{TempEqCnt}}

Consider a TDD-based massive MIMO system composed of $L$ hexagonal macro-cells, denoted by $\mathcal{L} =\text{ }\left\{ 1,\text{ }2,\text{ }.\text{ }.\text{ }.\text{ },\text{ }L \right\}$. Each macro-cell has a radius of ${{r}_{c}}$, where a BS is deployed in the center of each cell. Assume that $K$ MSs are randomly and uniformly distributed over each cell except for a central disk of radius ${{r}_{d}}$ [23]\footnote{The assumption that MSs are not located within the central disk of each cell is to ensure that MSs will not be too close to their serving BSs so that the far-field propagation model is valid.}. Each BS is equipped with $M$ antennas and each MS is equipped with a single antenna. The wireless channel is time-frequency flat over ${{T}_{c}}$ symbols (one coherence time) and ${{F}_{c}}$ sub-carriers (one coherence bandwidth), which is defined as one coherence block. In each cell, orthogonal pilot sequences are assigned to different MSs to avoid intra-cell interference, which occupies ${{F}_{c}} {{\tau }_{P}}$ time frequency resources ($0<{{\tau }_{P}}\le {{T}_{c}}$). In this paper, the number of simultaneously served MSs in one cell is assumed to be $K={{F}_{c}} {{\tau }_{P}}$ for the ease of analysis. The same set of pilot sequences are reused in different cells with shifted time resources [9]. Due to the limited time-frequency resources, it is difficult to ensure the non-overlapped pilot transmission of all cells. Therefore, the time shifted pilot transmission is conducted by cell groups like frequency reuse schemes [9]. First of all, the whole cell set $\mathcal{L} $ is partitioned into $\Gamma $ exclusive groups ${{A}_{1}},\text{ }{{A}_{2}},\text{ }\cdots \text{ },\text{ }{{A}_{\Gamma }}$, where $\Gamma ={{b}^{2}}+{{c}^{2}}+bc$, $b,c=0,1,2,\cdots $, and $b+c\ne 0$. The number of cells in cell group ${{A}_{i}}$ is denoted by $\left| {{A}_{i}} \right|$. Cells in the same group use the same time-frequency resources for UL pilot transmission. An example is shown in Fig. 1, illustrating the transmission of TSP with $\Gamma \text{=7}$. The transmission of each frame is with the length of ${{T}_{c}}$, which is composed of UL pilot transmission stage, cross-link (CL) data transmission stage, pure DL (PD) data transmission stage and UL data transmission stage. The frame of each group starts with its own first pilot symbol, which means that frames of different groups are not synchronous [20], [23]. The MS-BS channels corresponding to different frames are uncorrelated. As shown in the right side of Fig. 1, for each group of cells, channel estimation is conducted firstly in each frame and then be used to generate precoding/combing vector for DL/UL transmission. When the $i$-th group ${{A}_{i}}$ starts to transmit pilot in the $n$-th frame, ${{A}_{1}},\cdots ,{{A}_{i-1}}$ groups transmit DL data using the precoding vector based on the channel estimation of the $n$-th frame and ${{A}_{i+1}},\cdots ,{{A}_{\Gamma }}$ groups transmit DL data using the precoding vector based on the channel estimation of the $\left( n-1 \right)$-th frame. Hence, the UL pilot received at one BS is interfered by the UL pilots from the cells in the same group and the DL data from other groups. To ensure the non-overlapped pilot transmissions from different groups, $\Gamma -1\le {{T}_{d}}/{{\tau }_{P}}$, where ${{T}_{d}}$ is the length of DL data on one subcarrier and within each ${{T}_{c}}$. The length of UL data on one subcarrier and within each ${{T}_{c}}$ is denoted by ${{T}_{u}}$.

\setcounter{TempEqCnt}{\value{equation}}
\setcounter{equation}{1}
\begin{figure*}[ht]
\begin{equation}
{{\mathbf{\hat g}}_{llk'}}\! =\! {{\mathbf{g}}_{llk'}}\! +\! \underbrace {\sum\limits_{j \ne l,j \in {A_p}}^{} {\sqrt {\frac{{\rho _{UL,jk'}^P}}{{\rho _{UL,lk'}^P}}} {{\mathbf{g}}_{ljk'}}} }_{{{\mathbf{e}}_{llk',pilot}}}\! +\! \underbrace {\frac{{\left( {\sum\limits_{d = 1,d \notin {A_p}}^L {{{\mathbf{G}}_{ld}}\sum\limits_{k = 1}^K {\sqrt {\rho _{DL,dk}^D} {{\mathbf{w}}_{dk}}{\mathbf{x}}_{dk}^D} } } \right) \cdot {\mathbf{\psi }}_{k'}^H}}{{{F_c}{\tau _P}\sqrt {\rho _{UL,lk'}^P} }}}_{{{\mathbf{e}}_{llk',data}}} + \underbrace {\frac{{{{\mathbf{n}}_{P,l}} \cdot {\mathbf{\psi }}_{k'}^H}}{{{F_c}{\tau _P}\sqrt {\rho _{UL,lk'}^P} }}}_{{{\mathbf{e}}_{llk',noise}}},
\end{equation}
\end{figure*}

\begin{figure*}[ht]
\begin{equation}
{\varepsilon _{llk'}} \approx {\tilde \varepsilon _{llk'}} = \underbrace {\sum\limits_{j \ne l,j \in {A_p}}^{} {\frac{{\rho _{UL,jk'}^P}}{{\rho _{UL,lk'}^P}}{\beta _{ljk'}}} }_{{\varepsilon _{llk',pilot}}}{+}\underbrace {\frac{{P_{DL}^D}}{{{F_c}{\tau _P}\rho _{UL,lk'}^P}}\sum\limits_{d \notin {A_p}}^L {{\alpha _{ld}}} }_{{\varepsilon _{llk',data}}} + \underbrace {\frac{{\sigma _P^2}}{{{F_c} \cdot {\tau _P} \cdot \rho _{UL,lk'}^P}}}_{{\varepsilon _{llk',noise}}},
\end{equation}

\hrulefill
\vspace{-0.5cm}
\end{figure*}

\setcounter{equation}{\value{TempEqCnt}}

Let the $l$-th cell belong to the group ${{A}_{p}}$. During the UL pilot transmission of the group ${{A}_{p}}$, the pilot signal received at the BS of the $l$-th cell, i.e., the $l$-th BS, is given by (1), where $\mathbf{y}_{l}^{{}}$ is an $M\times {{F}_{c}}{{\tau }_{P}}$ matrix, $\rho _{UL,lk}^{P}\le \rho _{UL}^{P}$ is the UL pilot transmission power of $k$-th MS in the $l$-th cell, $\rho _{UL}^{P}$ is the largest pilot transmission power of MS, $\rho _{DL,dk}^{D}$ is DL data transmission power for $k$-th MS in the $l$-th cell, which satisfies $\sum\limits_{k=1}^{K}{\rho _{DL,dk}^{D}}=\rho _{DL}^{D}$, $\rho _{DL}^{D}$ is the total DL data transmission power of BS. ${{\mathbf{g}}_{ljk}}\in {{\mathbb{C}}^{M\times 1}}$ is the UL channel vector from the $k$-th MS in the $j$-th cell to the $l$-th BS, ${{\mathbf{\psi }}_{k}}\in {{\mathbb{C}}^{1\times {{F}_{c}} {{\tau }_{P}}}}$ denotes the mutually orthogonal pilot sequence allocated to the $k$-th MS with ${{\mathbf{\psi }}_{k}}\cdot \mathbf{\psi }_{k'}^{H}={{F}_{c}} {{\tau }_{P}} {{\delta }_{kk'}}$ [9], [20], [23], where ${{\delta }_{kk'}}$ is the Kronecker delta function, ${{\mathbf{G}}_{ld}}\in {{\mathbb{C}}^{M\times M}}$ is the channel matrix from the $d$-th BS to the $l$-th BS, ${{\mathbf{w}}_{dk}}\in {{\mathbb{C}}^{M\times 1}}$ is the normalized precoding vector for the $k$-th MS in the $d$-th cell, i.e., $\left\| {{\mathbf{w}}_{dk}} \right\|=1$, $\mathbf{x}_{dk}^{D}\in {{\mathbb{C}}^{1\times {{F}_{c}} {{\tau }_{P}}}}$ is the vectorized DL data for the $k$-th MS in the $d$-th cell and $\mathbf{n}_{P,l}^{{}}\sim \text{ }\mathcal{C}\mathcal{N}\left( \mathbf{0},\sigma _{P}^{2}{{\mathbf{I}}_{M {{F}_{c}} {{\tau }_{P}}}} \right)$ denotes the $M\times {{F}_{c}} {{\tau }_{P}}$ noise matrix in the $l$-th cell, where $\sigma _{P}^{2}$ is the noise variance during the pilot transmission stage. Given the channel vector ${{\mathbf{g}}_{ljk}}=\sqrt{{{\beta }_{ljk}}}\cdot {{\mathbf{h}}_{ljk}}$, where ${{\beta }_{ljk}}={{ {{d}_{ljk}^{-\eta }} }}{{\vartheta }_{ljk}}$ denotes the large scale fading. ${{d}_{ljk}}$ and ${{\vartheta }_{ljk}}$ are the distance and the shadow fading between the $k$-th MS in the $j$-th cell and the $l$-th BS, respectively, $\eta >2$ is the decay exponent, and ${{\mathbf{h}}_{ljk}}\sim \text{ }\mathcal{C}\mathcal{N}\left( \mathbf{0},{{\mathbf{I}}_{M}} \right)$ represents the $M\times 1$ small scale fading vector. The shadow fading ${{\vartheta }_{ljk}}$ is modeled via a log-normal distributed variable, i.e., $10{{\log }_{10}}\left( {{\vartheta }_{ljk}} \right)\sim \mathcal{N}\left( 0,\sigma _{\text{sh}}^{2} \right)$ , where ${{\sigma }_{sh}}$ is the logarithmic standard deviation [24]. Similarly, the channel between the $l$-th BS and the $d$-th BS is modeled as ${{\mathbf{G}}_{ld}}=\sqrt{{{\alpha }_{ld}}}{{\mathbf{D}}_{ld}}$, where ${{\alpha }_{ld}}={{ {{d}_{BS,ld}^{-\eta }} }}{{\vartheta }_{BS,ld}}$ is the large scale fading, ${{d}_{BS,ld}}$ and ${{\vartheta }_{BS,ld}}$ are the distance and the shadow fading between the $l$-th BS and the $d$-th BS, respectively, and ${{\mathbf{D}}_{ld}}$ is the $M\times M$ small scale fading matrix. Note that BSs of macro cells are usually installed at high places and line of sight (LOS) paths may exist between BSs. In addition, there is not enough local scattering around BS antennas, which leads to the strong spatial correlation [25]-[29]. Thus the small scale fading matrix  ${{\mathbf{D}}_{ld}}$ is modeled as a correlated Ricean one, i.e., ${{\mathbf{D}}_{ld}}=\frac{\sqrt{{{k}_{T}}}}{\sqrt{{1+}{{k}_{T}}}}{{\mathbf{\bar{C}}}_{ld}}+\frac{\text{1}}{\sqrt{{1+}{{k}_{T}}}}{{\mathbf{C}}_{ld}}$, where ${{k}_{T}}$ is the Ricean factor, $\frac{\sqrt{{{k}_{T}}}}{\sqrt{{1+}{{k}_{T}}}}{{\mathbf{\bar{C}}}_{ld}}$ accounts for the $M\times M$ LOS path component, ${{\mathbf{\bar{C}}}_{ld}}$ is the array response vector decided by the angle of departure and angle of arrival [30]-[31], and $\frac{\text{1}}{\sqrt{{1+}{{k}_{T}}}}{{\mathbf{C}}_{ld}}=\frac{\text{1}}{\sqrt{{1+}{{k}_{T}}}}\mathbf{R}_{R,ld}^{\frac{1}{2}}{{\mathbf{H}}_{W,ld}}\mathbf{R}_{T,ld}^{\frac{1}{2}}$ is the $M\times M$ correlated scattering component. ${{\mathbf{R}}_{R,ld}}$ and ${{\mathbf{R}}_{T,ld}}$ are the correlation matrices at the receiver and transmitter, respectively, ${{\mathbf{H}}_{W,ld}}$ is the independent Rayleigh channel matrix whose entries follow i.i.d complex Gaussian distribution, i.e., ${{\left[ {{\mathbf{H}}_{W,ld}} \right]}_{p,q}}\sim \mathcal{C}\mathcal{N}\left( 0,1 \right)$. Since all the BSs are assumed to be equipped with the same antenna configuration, ${{\mathbf{R}}_{R,ld}}={{\mathbf{R}}_{T,ld}}=\mathbf{R}$ for all $l$ and $d$. $\mathbf{R}$ is modeled via the widely-used exponential model of Loyka, i.e., ${{\left[ \mathbf{R} \right]}_{p,q}}={{\kappa }^{\left| p-q \right|}}$, where $\kappa \in \left[ 0,1 \right]$ is the adjacent antenna correlation coefficient (or spatial correlation coefficient) [25]-[26]. Thus, $\mathbf{R}$  is a real symmetric matrix, and the channel becomes more correlated when $\kappa $ gets larger.

Throughout this paper, the matched filtering (MF) method is used for precoding and detection due to its simplicity for analysis. Furthermore, the performances of other linear precoding and detection such as zero-forcing (ZF) method are evaluated by simulations, where the ZF method shows similar trend with the MF method.

\section{Performance of Channel Estimation}

\subsection{Analysis of Channel Estimation Error}
In the $l$-th cell ($l\in {{A}_{p}}$), the channels between the ${k}'$-th MS and the $l$-th BS can be estimated by ${{\mathbf{\hat g}}_{llk'}} = {{\left( {{{\mathbf{y}}_l}{\mathbf{\psi }}_{k'}^H} \right)} \mathord{\left/
 {\vphantom {{\left( {{{\mathbf{y}}_l}{\mathbf{\psi }}_{k'}^H} \right)} {\left( {{F_c}{\tau _P}\sqrt {\rho _{UL,lk'}^P} } \right)}}} \right.
 \kern-\nulldelimiterspace} {\left( {{F_c}{\tau _P}\sqrt {\rho _{UL,lk'}^P} } \right)}}$, which is further expanded by (2), where $\mathbf{e}_{llk'}^{{}}=\mathbf{e}_{llk',pilot}^{{}}+\mathbf{e}_{llk',data}^{{}}+\mathbf{e}_{llk',noise}^{{}}$ is the $M\times 1$ channel estimation error, composed of intra-group interference $\mathbf{e}_{llk',pilot}^{{}}$ caused by UL pilot transmission from cells in the same group ${{A}_{p}}$, inter-group interference $\mathbf{e}_{llk',data}^{{}}$ caused by DL data transmission from other groups, and background noise $\mathbf{e}_{llk',noise}^{{}}$. The MSCEE of the $k'$-th MS in the $l$-th cell is defined as ${{\varepsilon }_{llk'}}=\frac{1}{M}\mathbb{E}\left\{ {{\left\| \mathbf{e}_{llk'}^{{}} \right\|}^{2}} \right\}$. Omitting the weak correlation between the precoding vector ${{\mathbf{w}}_{dk}}$ and the BS-BS channel ${{\mathbf{G}}_{ld}}$, ${{\varepsilon }_{llk'}}$ is approximated by (3)(see Appendix A), where $\varepsilon _{llk',pilot}^{{}}$, $\varepsilon _{llk',data}^{{}}$, and $\varepsilon _{llk',noise}^{{}}$ stand for the impact of intra-group pilot interference, inter-group data interference, and noise, respectively. It can be seen that the MSCEE is independent of the spatial correlation coefficient $\kappa $ and the Rician factor $k_T$. This is because the spatial correlation does not impact the total power of interference. For $\Gamma =1$, the TSP is equivalent to the AP and there is only intra-group pilot interference, so ${{\varepsilon }_{llk'}}={{E}_{\vartheta }}\sum\limits_{j\ne l}^{{L}}{\frac{\rho _{UL,jk'}^{P}}{\rho _{UL,lk'}^{P}}{{{{d}_{ljk'}^{-\eta }}}}}+\frac{\sigma _{P}^{2}}{{{F}_{c}} {{\tau }_{P}} \rho _{UL,lk'}^{P}}$ (exact result). When $\Gamma $ increases from 1 to 3, ${{\varepsilon }_{llk'}}$ increases significantly, because $\frac{2}{3}L$ cells (including the nearest 6 cells) generate high powered DL data interference instead of relatively low powered UL pilot interference.

\begin{figure}[h]
  \centering
  \includegraphics[angle=0,width=0.5\textwidth]{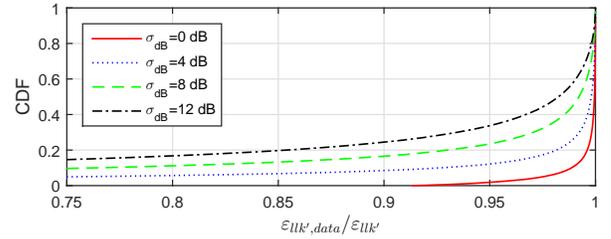}\\
  \caption{The CDF of $\frac{{{\varepsilon }_{llk',data}}}{{{\varepsilon }_{llk'}}}$.}\label{Fig. 2}
\end{figure}

\setcounter{TempEqCnt}{\value{equation}}
\setcounter{equation}{3}
\begin{figure*}[ht]
\begin{equation}
\begin{array}{*{20}{l}}
  {{\mathbf{y}}_{lk'}^{UL}}&\!\!{ = {{\mathbf{a}}_{lk'}}\left( {\sum\limits_{k = 1}^K {\sqrt {\rho _{UL,lk}^D} {\mathbf{g}}_{llk}^{}{\mathbf{x}}_{lk}^U}  + \sum\limits_{j = 1,j \ne l}^L {\sum\limits_{k = 1}^K {\sqrt {\rho _{UL,jk}^D} {\mathbf{g}}_{ljk}^{}{\mathbf{x}}_{jk}^U} }  + {\mathbf{n}}_{UL,lk'}^{}} \right)} \\
  {}&\!\!{ = \underbrace {\sqrt {\rho _{UL,lk'}^D} {{\mathbf{a}}_{lk'}}{\mathbf{g}}_{llk'}^{}{\mathbf{x}}_{lk'}^U}_{{\text{target signal}}} +\!\! \underbrace {\sum\limits_{k = 1,k \ne k'}^K {\sqrt {\rho _{UL,lk}^D} {{\mathbf{a}}_{lk'}}{\mathbf{g}}_{llk}^{}{\mathbf{x}}_{lk}^U} }_{{\text{intra-cell interference}}} +\!\! \underbrace {\sum\limits_{j \in {A_p},j \ne l}^{} {\sum\limits_{k = 1}^K {\sqrt {\rho _{UL,jk}^D} {{\mathbf{a}}_{lk'}}{\mathbf{g}}_{ljk}^{}{\mathbf{x}}_{jk}^U} } }_{{\text{intra-group interference}}}} \\
  {}&{ \quad + \underbrace {\sum\limits_{j=1, j \notin {A_p}}^L {\sum\limits_{k = 1}^K {\sqrt {\rho _{UL,jk}^D} {{\mathbf{a}}_{lk'}}{\mathbf{g}}_{ljk}^{}{\mathbf{x}}_{jk}^U} } }_{{\text{inter-group interference}}} + \underbrace {{{\mathbf{a}}_{lk'}}{\mathbf{n}}_{UL,lk'}^{}}_{{\text{noise}}}} ,
\end{array}
\end{equation}
\end{figure*}

\setcounter{equation}{\value{TempEqCnt}}

\setcounter{TempEqCnt}{\value{equation}}
\setcounter{equation}{5}
\begin{figure*}[ht]
\begin{equation}
\begin{gathered}
  {\mathbf{y}}_{lk'}^{CL} \!= \!\underbrace {\sqrt {\rho _{DL,lk'}^D} {\mathbf{g}}_{llk'}^T{{\mathbf{w}}_{lk'}}{\mathbf{x}}_{lk'}^D}_{{\text{target signal}}} + \underbrace {{\mathbf{g}}_{llk'}^T\!\!\sum\limits_{k = 1,k \ne k'}^K \!\!{\sqrt {\rho _{DL,lk}^D} {{\mathbf{w}}_{lk}}{\mathbf{x}}_{lk}^D} }_{{\text{intra-cell interference}}} +\!\! \underbrace {\sum\limits_{j \ne l,j \in {A_p}}^{}\!\! {{\mathbf{g}}_{jlk'}^T\sum\limits_{k = 1}^K {\sqrt {\rho _{DL,jk}^D} {{\mathbf{w}}_{jk}}{\mathbf{x}}_{jk}^D} } }_{{\text{intra-group interference}}} \\
   + \underbrace {\sum\limits_{j \in {A_q}}^{} {\sum\limits_{k = 1}^K {\sqrt {\rho _{UL,jk}^P} {g_{lk'jk}}{{\mathbf{\psi }}_k}} } }_{{\text{inter-group UL pilot interference}}} + \underbrace {\sum\limits_{j=1, j \notin {A_p},j \notin {A_q}}^L {{\mathbf{g}}_{jlk'}^T\sum\limits_{k = 1}^K {\sqrt {\rho _{DL,jk}^D} {{\mathbf{w}}_{jk}}{\mathbf{x}}_{jk}^D} } }_{{\text{inter-group interference}}} + \underbrace {{\mathbf{n}}_{DL-CL,lk'}^{}}_{{\text{noise}}}, \\
\end{gathered}
\end{equation}

\hrulefill
\vspace{-0.5cm}
\end{figure*}

\setcounter{equation}{\value{TempEqCnt}}

In (3), the MSCEE is related to the large scale fading between the BS of target cell and MSs using the same pilot with target MS (${{\beta }_{ljk'}}$), and the large scale fading between the BS of target cell and the BSs of interfering cells (${{\alpha }_{ld}}$). As seen from (3), the transmission power of the interferer in $\varepsilon _{llk',data}^{{}}$, i.e., $\frac{P_{DL}^{D}}{{{F}_{c}}{{\tau }_{P}}}=\frac{P_{DL}^{D}}{K}$, is much larger than the transmission power in $\varepsilon _{llk',pilot}^{{}}$, i.e., $\rho _{UL,jk'}^{P}$. The distance between the interferer and the target BS in $\varepsilon _{llk',data}^{{}}$ is also smaller than that in $\varepsilon _{llk',pilot}^{{}}$. Therefore, the inter-group data interference $\varepsilon _{llk',data}^{{}}$ is expected to be the dominant composition of the MSCEE. However, the large scale fading is affected by both the instantaneous location of MSs and the shadow fading, while the randomness of the shadow fading makes the MSCEE and the composition of the MSCEE being fluctuating. Therefore, to validate the hypothesis that the MSCEE is dominated by the inter-group data interference, we plot the numerical cumulative distribution function (CDF) of $\frac{{{\varepsilon }_{llk',data}}}{{{\varepsilon }_{llk'}}}$. Fig. 2 shows the CDF of $\frac{{{\varepsilon }_{llk',data}}}{{{\varepsilon }_{llk'}}}$ with different logarithmic standard deviations of shadow fading (i.e., ${{\sigma }_{sh}}$) where 10000 random realizations of user locations and shadow fading profiles are generated. In this simulation, the group number is chosen to be $\Gamma =7$ and other system parameters are listed in Table I (at the beginning of Sec. V). It is shown that $\frac{{{\varepsilon }_{llk',data}}}{{{\varepsilon }_{llk'}}}$ is higher than 85\% for at least 80\% of samples. With the decrease of ${{\sigma }_{sh}}$, the dominance of the inter-group data interference in the MSCEE is strengthened since the randomness of $\frac{{{\varepsilon }_{llk',data}}}{{{\varepsilon }_{llk'}}}$ is getting weaker. Therefore, it is clear that the MSCEE is dominated by the inter-group data interference.

\subsection{The impact of the MSCEE on the SINR}
At the UL data transmission stage, the detected signal of the $k'$-th MS in the $l$-th cell ($l\in {{A}_{p}}$) at its serving BS is given by (4), where ${{\mathbf{a}}_{lk'}}=\mathbf{\hat{g}}_{llk'}^{H}$ is the MF detection vector [1, 9, 32] for the $k'$-th MS in the $l$-th cell, $\rho _{UL,lk}^{D}\le \rho _{UL}^{D}$ is the UL data transmission power of $k$-th MS in the $l$-th cell, $\rho _{UL}^{D}$ is the largest UL data transmission power of MS, $\mathbf{x}_{lk}^{U}\in {{C}^{1\times {{F}_{c}}\cdot {{T}_{u}}}}$ is the UL data of the $k$-th MS in the $l$-th cell, $\mathbf{n}_{UL,lk'}^{{}}\sim \mathcal{C}\mathcal{N}(\mathbf{0},\sigma _{UL}^{2}{{\mathbf{I}}_{M\cdot {{F}_{c}}\cdot {{T}_{u}}}})$ is the $M\times {{F}_{c}}\cdot {{T}_{u}}$ additive Gaussian noise matrix and ${{\sigma }_{UL}}^{2}$ is noise variance for UL data transmission stage. The intra-group interference $\sum\limits_{j\in {{A}_{p}},j\ne l}^{L}{\sum\limits_{k=1}^{K}{\sqrt{\rho _{UL,jk}^{D}}{{\mathbf{a}}_{lk'}}\mathbf{g}_{ljk}^{{}}\mathbf{x}_{jk}^{U}}}=\underbrace{\sum\limits_{j\in {{A}_{p}},j\ne l}^{L}{\sqrt{\rho _{UL,jk'}^{D}}\left( \mathbf{g}_{ljk'}^{H} \right)\mathbf{g}_{ljk'}^{{}}\mathbf{x}_{jk'}^{U}}}_{\text{correlated  interference }}+$ $\underbrace{\sum\limits_{j\in {{A}_{p}},j\ne l}^{L}{\sqrt{\rho _{UL,jk'}^{D}}{{\left( \mathbf{\hat{g}}_{llk'}^{{}}-\mathbf{g}_{ljk'}^{{}} \right)}^{H}}\mathbf{g}_{ljk'}^{{}}\mathbf{x}_{jk'}^{U}}}_{\text{uncorrelated  interference }}+$ $\underbrace{\sum\limits_{j\in {{A}_{p}},j\ne l}^{L}{\sum\limits_{k=1,k\ne k'}^{K}{\sqrt{\rho _{UL,jk}^{D}}{{\mathbf{a}}_{lk'}}\mathbf{g}_{ljk}^{{}}\mathbf{x}_{jk}^{U}}}}_{\text{uncorrelated  interference }}$, where the power of correlated interference is proportional to ${{M}^{2}}$ while the power of uncorrelated interference is only proportional to $M$. This can be proved in the process of deriving the UL SINR in Appendix B. Note that all other intra-cell UL data interference, intra-group UL data interference and inter-group UL data interference are uncorrelated interferences.

Utilizing the properties of Chi-square distribution, a closed form UL SINR is obtained as (see Appendix B)

\setcounter{equation}{4}
\begin{equation}
\!S\!I\!N\!R_{lk'}^{UL}\! =\! \frac{{\left( {M + 1} \right)\beta _{llk'}^2 + \varepsilon _{llk'}^{}\beta _{llk'}^{}}}{{M\!\!\!\!\!\sum\limits_{j \ne l,j \in {A_p}}^{}\!\!\! {\frac{{\rho _{UL,jk'}^D}}{{\rho _{UL,lk'}^D}}\frac{{\rho _{UL,jk'}^P}}{{\rho _{UL,lk'}^P}}\beta _{ljk'}^2}\!  +\! \left( {\beta _{llk'}^{}\! +\! \varepsilon _{llk'}^{}} \!\right)\!{\varsigma _{UL,lk'}}}},
\end{equation}
where $M \!\!\!\sum\limits_{j\ne l,j\in {{A}_{p}}}^{{}}{\frac{\rho _{UL,j{k}'}^{D}}{\rho _{UL,l{k}'}^{D}}\frac{\rho _{UL,jk'}^{P}}{\rho _{UL,lk'}^{P}}\beta _{lj{k}'}^{2}}$ in the denominator shows the impact of correlated intra-group data interference, $\left( \beta _{ll{k}'}^{{}}\!+\!\varepsilon _{ll{k}'}^{{}} \right){{\varsigma }_{UL,l{k}'}}$ shows the impact of all uncorrelated interference plus noise, and ${{\varsigma }_{UL,l{k}'}}=\sum\limits_{j=1}^{L}{\sum\limits_{k=1}^{K}{\frac{\rho _{UL,jk}^{D}}{\rho _{UL,l{k}'}^{D}}\beta _{ljk}^{{}}}}-\beta _{llk'}^{{}}+\frac{\sigma _{UL}^{2}}{\rho _{UL,l{k}'}^{D}}$.

\setcounter{TempEqCnt}{\value{equation}}
\setcounter{equation}{6}
\begin{figure*}[ht]
\begin{equation}
SINR_{lk'}^{CL} \approx \overline {SINR} _{lk'}^{CL} = \frac{{\left( {M + 1} \right)\beta _{llk'}^2 + \varepsilon _{llk'}^{}\beta _{llk'}^{}}}{{M\sum\limits_{j \ne l,j \in {A_p}}^{} {\frac{{\beta _{llk'}^{} + {\varepsilon _{llk'}}}}{{\beta _{jjk'}^{} + {\varepsilon _{jjk'}}}}\frac{{\rho _{DL,jk'}^D}}{{\rho _{DL,lk'}^D}}\frac{{\rho _{UL,lk'}^P}}{{\rho _{UL,jk'}^P}}\beta _{jlk'}^2} {+}\left( {\beta _{llk'}^{} + {\varepsilon _{llk'}}} \right){\varsigma _{CL,lk'}}}},
\end{equation}

\end{figure*}

\setcounter{equation}{\value{TempEqCnt}}

\setcounter{TempEqCnt}{\value{equation}}
\setcounter{equation}{7}
\begin{figure*}[ht]
\begin{equation}
SINR_{lk'}^{PD} \approx \overline {SINR} _{lk'}^{PD} = \frac{{\left( {M + 1} \right)\beta _{llk'}^2 + \varepsilon _{llk'}^{}\beta _{llk'}^{}}}{{M\sum\limits_{j \ne l,j \in {A_p}}^{} {\frac{{\beta _{llk'}^{} + {\varepsilon _{llk'}}}}{{\beta _{jjk'}^{} + {\varepsilon _{jjk'}}}}\frac{{\rho _{DL,jk'}^D}}{{\rho _{DL,lk'}^D}}\frac{{\rho _{UL,lk'}^P}}{{\rho _{UL,jk'}^P}}\beta _{jlk'}^2} {+}\left( {\beta _{llk'}^{} + {\varepsilon _{llk'}}} \right){\varsigma _{PD,lk'}}}},
\end{equation}

\hrulefill
\vspace{-0.5cm}
\end{figure*}

\setcounter{equation}{\value{TempEqCnt}}

Now we move to the DL transmission, as shown in Fig. 1, which is divided into two stages, i.e., a CL stage when both UL pilot and DL data transmission happen and a PD stage when only DL data transmission occur. MF precoding is employed for DL data transmission, i.e., ${{\bf{w}}_{lk}}{\rm{ = }}\frac{{{{{\bf{\hat g}}}_{llk}}^*}}{{\left\| {{{{\bf{\hat g}}}_{llk}}} \right\|}}$ [1], [9], 20]. At the CL stage, when the cell group $A_p$ is at the DL data transmission mode and the cell foup $A_q$ is at the UL pilot transmission mode, the signal received at the $k'$-th MS in the $l$-th cell (for example, $l\in {{A}_{p}}={{A}_{1}}$ and ${{A}_{1}}=\left\{ 1,\text{ }21,\text{ }24,\text{ 27},\text{ }30,\text{ 33},\text{ }36 \right\}$ in Fig. 1) is given by (6), where ${{g}_{lk'jk}}=\sqrt{{{\mu }_{lk'jk}}}{{\gamma }_{lk'jk}}$ is the channel between the $k$-th MS in the $j$-th cell and the $k'$-th MS in the $l$-th cell, ${{\mu }_{lk'jk}}=d_{lk'jk}^{-\eta }{{\vartheta }_{lk'jk}}$ is the large scale fading modeled similarly with that of channel between MS and BS, and ${{\gamma }_{lk'jk}}\sim \text{ }\mathcal{C}\mathcal{N}\left( 0,1 \right)$ is the small scale fading, , and $\mathbf{n}_{DL-CL,dk'}^{{}}\sim \mathcal{C}\mathcal{N}\left( \mathbf{0},\sigma _{CL}^{2}{{\mathbf{I}}_{1\times {{F}_{c}}{{\tau }_{P}}}} \right)$ is the $1\times {{F}_{c}}{{\tau }_{P}}$ background noise, $\sigma _{CL}^{2}$ is the noise variance for the CL stage. Similar to the UL, the intra-group DL data interference is also composed of correlated DL data interference caused by pilot reusing in the same group and the rest uncorrelated DL data interference.

A closed form DL SINR at the CL stage is approximated as (7)(see Appendix C), where $M\sum\limits_{j\ne l,j\in {{A}_{p}}}^{{}}{\frac{\beta _{llk'}^{{}}+{{\varepsilon }_{llk'}}}{\beta _{jjk'}^{{}}+{{\varepsilon }_{jjk'}}}\frac{\rho _{DL,jk'}^{D}}{\rho _{DL,lk'}^{D}}\frac{\rho _{UL,lk'}^{P}}{\rho _{UL,jk'}^{P}}\beta _{jlk'}^{2}}$ shows the impact of correlated intra-group data interference and $\left( \beta _{llk'}^{{}}\!+\!{{\varepsilon }_{llk'}} \right){{\varsigma }_{CL,lk'}}$ shows the impact of all uncorrelated interference plus noise with ${{\varsigma }_{CL,l{k}'}}\!=\!\frac{\rho _{DL}^{D}}{\rho _{DL,l{k}'}^{D}}\!\!\sum\limits_{j=1,j\notin {{A}_{q}}}^{L}\!\!{\beta _{jl{k}'}^{{}}}-\beta _{ll{k}'}^{{}}+\sum\limits_{j\in {{A}_{q}}}^{{}}{\sum\limits_{k=1}^{K}{\frac{\rho _{UL,jk}^{P}}{\rho _{DL,l{k}'}^{D}}\mu _{l{k}'jk}^{{}}}}+\frac{\sigma _{CL}^{2}}{\rho _{DL,l{k}'}^{D}}$.

At the PD data transmission stage, the target MS is only interfered by the DL data of MSs from all other cells. Using the same MF precoding, the corresponding SINR of PD data transmission stage can be approximated similarly in (8), where ${{\varsigma }_{PD,lk'}}=\frac{\rho _{DL}^{D}}{\rho _{DL,lk'}^{D}}\sum\limits_{j=1}^{L}{\beta _{jlk'}^{{}}}-\beta _{llk'}^{{}}+\frac{\sigma _{PD}^{2}}{\rho _{DL,lk'}^{D}}$, and $\sigma _{PD}^{2}$ is the power of noise for PD data transmission stage.

Compared to the UL SINR in (5), the first term in the denominator of $\overline{SINR}_{lk'}^{CL}$ and $\overline{SINR}_{lk'}^{PD}$ are scaled by $\frac{\beta _{llk'}^{{}}+{{\varepsilon }_{llk'}}}{\beta _{jjk'}^{{}}+{{\varepsilon }_{jjk'}}}$ since the precoding vector for DL transmission is normalized for each MS. Due to the different large scale fading and power allocation, the impact of this scaling factor is different from one MS to another. However, from the statistical point of view (i.e., when considering the average SINR performance over large numbers of random realizations), the impact of this scaling factor will average out since $\mathbb{E}\left\{ \frac{\beta _{llk'}^{{}}+{{\varepsilon }_{llk'}}}{\beta _{jjk'}^{{}}+{{\varepsilon }_{jjk'}}} \right\}\approx 1$. Therefore, the analysis in the following will be derived using $SINR_{lk'}^{UL}$ considering UL transmission, which are expected to be also valid for the average SINR performance of CL transmission and PD transmission over large numbers of random realizations.

Seen from (5), (7) and (8), the SINRs are affected by the transmission power of (for) all MSs in the whole system. When the path-loss based power control is considered, the transmission power of (for) each MS is proportional to the path-loss of the channel between the MS and its serving BS, i.e., $\rho _{UL,lk}^{P}=\frac{\beta _{llk}^{-1}}{\underset{v\in \left\{ 1,...,K \right\}}{\mathop{\max }}\,\left( \beta _{llv}^{-1} \right)}\rho _{UL}^{P}$, $\rho _{UL,lk}^{D}=\frac{\beta _{llk}^{-1}}{\underset{v\in \left\{ 1,...,K \right\}}{\mathop{\max }}\,\left( \beta _{llv}^{-1} \right)}\rho _{UL}^{D}$ and $\rho _{DL,dk}^{D}=\frac{\beta _{ddk}^{-1}}{\sum\limits_{v=1}^{K}{\beta _{ddv}^{-1}}}\rho _{DL}^{D}$. In this way, the average power of target signal of all MSs will be the same and the system achieves the best fairness performance. However, substituting these power settings into (5), (7) and (8), it can be found that no further insights can be derived. To show the impact of MSCEE on the SINRs of TSP with finite $M$, we will simplify the analysis by using uniform power allocation. In addition, the SINR performance using the path-loss based power control will be evaluated by simulations.

When uniform power allocation is used, the power allocation is given by $\rho _{UL,lk}^{P}=\rho _{UL}^{P}$, $\rho _{UL,lk}^{D}=\rho _{UL}^{D}$ and $\rho _{DL,dk}^{D}=\frac{1}{K}\rho _{DL}^{D}$, respectively. From the partial derivatives of $SINR_{lk'}^{UL}$, it can be proved that $SINR_{lk'}^{UL}$ is a decreasing function of $\varepsilon _{llk'}^{{}}$ and an increasing function of $M$ (which is expected to be valid for CL and PD stage and verified by simulations in Sec. V). The asymptotic performance over $M$ is given by
\setcounter{equation}{8}
\begin{equation}
\mathop {\lim }\limits_{M \to \infty } SINR_{lk'}^{UL} = \beta _{llk'}^2/\sum\limits_{j \ne l,j \in {A_p}}^{} {\beta _{ljk'}^2} .
\end{equation}
When $M$ grows to infinity, TSP achieves an ideal SINR performance that the impacts of MSCEE, the uncorrelated intra-group interference and the inter-group interference on the UL SINR become negligible. However, when $M$ is finite, the impact of MSCEE on the UL SINR is significant, which will be illustrated in the following.

To achieve a practical target $SIN{{R}_{\Upsilon }}$, the number of BS antennas needed for the TSP can be derived by solving $SINR_{l{k}'}^{UL}=SIN{{R}_{\Upsilon }}$, which is given by
\begin{equation}
\begin{array}{*{20}{l}}
  {{M_T}}&{ = \frac{{SIN{R_\Upsilon }\left( {\beta _{llk'}^{} + \varepsilon _{llk'}^{}} \right){\varsigma _{UL,lk'}} - \beta _{llk'}^{}\left( {\varepsilon _{llk'}^{} + \beta _{llk'}^{}} \right)}}{{\left( {\beta _{llk'}^2 - SIN{R_\Upsilon }\sum\limits_{j \ne l,j \in {A_p}}^{} {\beta _{ljk'}^2} } \right)}}} \\
  {}&{ > \left( {\beta _{llk'}^{} + \varepsilon _{llk'}^{}} \right)\frac{1}{{\beta _{llk'}^{}}}\left( {SIN{R_\Upsilon }\frac{{{\varsigma _{UL,lk'}}}}{{\beta _{llk'}^{}}} - 1} \right)}.
\end{array}
\end{equation}
It can be seen that ${{M}_{T}}$ increases rapidly with $\varepsilon _{ll{k}'}^{{}}$ with a slope larger than $\frac{1}{\beta _{ll{k}'}^{{}}}\left( SIN{{R}_{\Upsilon }}\frac{{{\varsigma }_{UL,l{k}'}}}{\beta _{ll{k}'}^{{}}}-1 \right)$, which is large and scales with $\frac{1}{\beta _{ll{k}'}^{{}}}$. Hence, with a small number of BS antennas, it is important to reduce the MSCEE in order to achieve a target performance.

\section{IC Based Time-shifted Pilot Scheme}
\begin{figure*}[htbp]
  \centering
  \includegraphics[angle=0,width=1\textwidth]{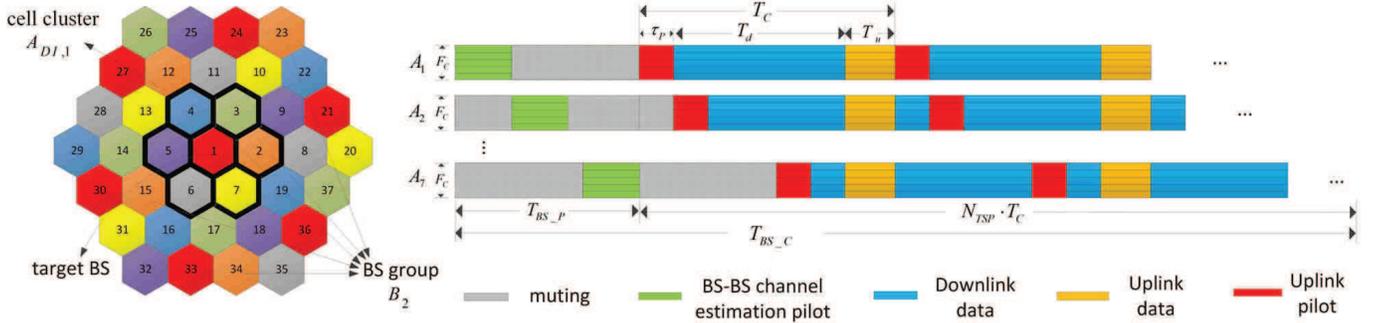}\\
  \caption{Illustration of the IC-TSP.}\label{Fig. 3}
\end{figure*}

As illustrated before, the channel estimation is severely contaminated by the inter-group interference from DL data transmission in other groups. Therefore, it is highly desirable to cancel out the inter-group interference. Note that the inter-group interference can be estimated using the DL data and precoding vectors shared among BSs. Although in distributed radio access networks (D-RAN), this data sharing requires a large backhaul among BSs, in centralized radio access networks (C-RAN) [35]-[38] and open radio access networks (O-RAN) [39], it can be naturally supported without much additional cost. With the idea to cancel out inter-group interference, an IC based channel estimation is proposed.

To cancel the dominant inter-group data interference ${{\mathbf{e}}_{llk',data}}$ in (2), the channel between the target BS and its main interfering BSs should be estimated. As shown in the right side of Fig. 3, this can be realized via a super TSP frame structure with the length of one coherence time of BS-BS channel ${{T}_{BS\_C}}$. The super TSP frame structure consists of two parts, i.e., the BS-BS channel estimation stage with a duration of ${{T}_{BS\_P}}$ at the beginning of each frame and ${{N}_{TSP}}$ consecutive TSP frames. Compared to MSs, BSs lack of mobility so it is expected that the coherence time of the BS-BS channel is much longer than that of BS-MS channels, i.e., ${{T}_{BS\_C}}\gg{{T}_{c}}$ and ${{N}_{TSP}}\gg1$. Assuming that ${{N}_{L}}\ge 1$ layers of BS-BS interference is to be canceled, channels between the target BS and up to ${{L}_{D\_main}}{=3}{{N}_{L}}\left( {{N}_{L}}+1 \right)\ge 6$ nearest BSs should be estimated during the BS-BS channel estimation stage, which is conducted in a round-robin manner. Define a cell cluster ${{A}_{DI,l}}$ consisting of the target cell $l$ and its ${{L}_{D\_main}}$ nearest cells. As shown in the left side of Fig. 3, considering ${{L}_{D\_main}}=6$, BSs in the cluster ${{A}_{DI,l}}=\left\{ 1,2,\cdots ,7 \right\}$ transmit pilot signals sequentially. The BS-BS channel estimation is also conducted like frequency reuse schemes with the reuse factor of ${{L}_{D\_main}}+1$. Thus, the BS-BS channel can be estimated without severe interference from nearby cells in the cluster.

\setcounter{TempEqCnt}{\value{equation}}
\setcounter{equation}{15}
\begin{figure*}[ht]
\begin{equation}
\begin{array}{*{20}{l}}
  {\varepsilon _{llk'}^{IC}}&{{\!\!\!\!=}\frac{1}{M}\mathbb{E}\left\{ {{{\left\| {{\mathbf{\hat g}}_{llk'}^{IC} - {\mathbf{g}}_{llk'}^{}} \right\|}^2}} \right\}} \\
  {}&{\!\!\! \approx \underbrace {\sum\limits_{j \ne l,j \in {A_p}}^{} {\frac{{\rho _{UL,jk'}^P}}{{\rho _{UL,lk'}^P}}{\beta _{ljk'}}} }_{{\varepsilon _{llk',pilot}}}{+}\underbrace {\frac{{P_{DL}^D}}{{{F_c}{\tau _P}\rho _{UL,lk'}^P}}\sum\limits_{d = 1,d \notin {A_p},d \notin {A_{DI,l}}}^L {{\alpha _{BS,ld}}} }_{{\varepsilon _{llk',data,others}}}{+}} \\
  {}&{\underbrace {\frac{{P_{DL}^D}}{{{F_c}{\tau _P}\rho _{UL,lk'}^P}}\sum\limits_{d \notin {A_p},d \in {A_{DI,l}}}^{} {\sum\limits_{b \ne d,b \in {B_d}}^{} {{\alpha _{BS,lb}}} } }_{\varepsilon _{llk',data,residual}^{IC}} + \underbrace {{L_{D\_main}}\frac{{\rho _{DL}^D}}{{\rho _{}^{BS - P}}}\frac{{\sigma _{BS}^2}}{{{F_c}  {\tau _P}\rho _{UL,lk'}^P}}}_{\varepsilon _{llk',noise,residual}^{IC}}{+}\underbrace {\frac{{\sigma _P^2}}{{{F_c} \cdot {\tau _P}  \rho _{UL,lk'}^P}}}_{{\varepsilon _{llk',noise}}}},
\end{array}
\end{equation}

\hrulefill
\vspace{-0.5cm}
\end{figure*}

\setcounter{equation}{\value{TempEqCnt}}

During the BS-BS channel estimation stage, the pilot signal received at the $l$-th BS from the $d$-th BS ($d\in {{A}_{DI,l}}$) is given by
\begin{equation}
\mathbf{Y}_{ld}^{BS}\!=\!\sqrt{\rho _{{}}^{BS-P}}{{\mathbf{G}}_{ld}}\mathbf{P}\!+\!\underbrace{\sqrt{\rho _{{}}^{BS-P}}\!\!\!\!\sum\limits_{b\ne d,b\in {{B}_{d}}}^{{}}\!\!\!\!{{{\mathbf{G}}_{lb}}\mathbf{P}}\!+\!\mathbf{N}_{ld}^{BS}}_{{{\mathbf{J}}_{ld}}},
\end{equation}
where $\rho _{{}}^{BS-P}$ is the pilot power for BS channel estimation, $\mathbf{P}\in {{\mathbb{C}}^{M\times {{\tau }_{BS}}}}$ is the pilot matrix, ${{\tau }_{BS}}$ is the length of pilot sequence on each BS antenna, ${{B}_{d}}$ denotes a group of BSs which transmit pilot signals simultaneously with the $d$-th BS (including the $d$-th BS). As shown in Fig. 3, ${{B}_{2}}=\left\{ 2,12,15,23,34 \right\}$. Here, $\mathbf{N}_{ld}^{BS}\sim \text{ }\mathcal{C}\mathcal{N}\left( \mathbf{0},\sigma _{BS}^{2}{{\mathbf{I}}_{{{M}^{2}}}} \right)$ is the $M\times {{\tau }_{BS}}$ Gaussian additive noise matrix. ${{\mathbf{J}}_{ld}}=\sqrt{\rho _{{}}^{BS-P}}\sum\limits_{b\ne d,b\in {{B}_{d}}}^{{}}{{{\mathbf{G}}_{lb}}\mathbf{P}}+\mathbf{N}_{ld}^{BS}$ is the sum of the interference and the noise.

Firstly, we consider the traditional LS BS-BS channel estimation, i.e., ${{\tau }_{BS}}=M$ and the pilot matrix satisfies $\frac{1}{M}\mathbf{P}\cdot {{\mathbf{P}}^{H}}={{\mathbf{I}}_{M}}$. In this way, the estimation of channel matrix from the $d$-th BS to the $l$-th BS is given by
\begin{equation}
{{\mathbf{\hat G}}_{ld}}\! =\! \frac{{{\mathbf{y}}_{ld}^{BS}{{\mathbf{P}}^H}}}{{M\sqrt {\rho _{}^{BS - P}} }} \!=\! {{\mathbf{G}}_{ld}} +\!\!\! \underbrace {\sum\limits_{b \ne d,b \in {B_d}}^{}\!\!\! {{{\mathbf{G}}_{lb}}} \! +\! \frac{{{\mathbf{n}}_{ld}^{BS}{{\mathbf{P}}^H}}}{{M\sqrt {\rho _{}^{BS - P}} }}}_{{{\mathbf{E}}_{ld}}},
\end{equation}
where ${{\mathbf{E}}_{ld}}$ denotes the BS-BS channel estimation error.

Given the estimated BS-BS channel ${{\mathbf{\hat{G}}}_{ld}}$, the target BS can estimate the main inter-group interference generated by DL data transmission of BSs in ${{A}_{DI,l}}$. Assuming that the DL data and precoding vectors are shared among the BSs in ${{A}_{DI,l}}$ , the estimated inter-group interference is given by
\begin{equation}
ICI_l^{}{{ = }}\sum\limits_{d \in {A_{DI,l}}, d \notin {A_p}} \!\!\!{{{{\mathbf{\hat G}}}_{ld}}\sum\limits_{k = 1}^K {\sqrt {\rho _{DL,lk}^D} {{\mathbf{w}}_{dk}}{\mathbf{x}}_{dk}^D} },
\end{equation}
which can then be canceled from the received signal as
\begin{equation}
\mathbf{\bar{y}}_{l}^{{}}=\mathbf{y}_{l}^{{}}-ICI_{l}^{{}}.
\end{equation}

In (1), the inter-group data interference in $\mathbf{y}_{l}^{{}}$ can be divided into two parts, i.e., the interference  from  cells   in  ${{A}_{DI,l}}$  and other  cells,  given  by $\sum\limits_{d\in {{A}_{DI,l}}, d\notin {{A}_{p}}}{{{\mathbf{G}}_{ld}}\sum\limits_{k=1}^{K}{\sqrt{\rho _{DL,dk}^{D}}{{\mathbf{w}}_{dk}}\mathbf{x}_{dk}^{D}}+}$
$\sum\limits_{d=1,d\notin {{A}_{p}},d\notin {{A}_{DI,l}}}^{L}\!{{{\mathbf{G}}_{ld}}\cdot}$
$({\sum\limits_{k=1}^{K}{\sqrt{\rho _{DL,dk}^{D}}{{\mathbf{w}}_{dk}}\mathbf{x}_{dk}^{D}}})$. The $ICI_{l}^{{}}$ in (13) can also be divided into two parts, given by $ICI_l^{} = \sum\limits_{d \in {A_{DI,l}}, d \notin {A_p}} {{{\mathbf{G}}_{ld}}\sum\limits_{k = 1}^K {\sqrt {\rho _{DL,lk}^D} {{\mathbf{w}}_{dk}}{\mathbf{x}}_{dk}^D} }  + \sum\limits_{d \in {A_{DI,l}}, d \notin {A_p}} {{{\mathbf{E}}_{ld}}\sum\limits_{k = 1}^K {\sqrt {\rho _{DL,lk}^D} {{\mathbf{w}}_{dk}}{\mathbf{x}}_{dk}^D} } $. The first part of inter-group data interference and $ICI_{l}^{{}}$ is identical. Therefore, after IC in (14), The residual inter-group data interference is given by $\sum\limits_{d=1,d\notin {{A}_{p}},d\notin {{A}_{DI,l}}}^{L}\!\!\!\!\!\!{{{\mathbf{G}}_{ld}}\sum\limits_{k=1}^{K}\!\!{\sqrt{\rho _{DL,dk}^{D}}{{\mathbf{w}}_{dk}}\mathbf{x}_{dk}^{D}}}-\!\!\! \sum\limits_{d \in {A_{DI,l}}, d \notin {A_p}} {{{\mathbf{E}}_{ld}}\sum\limits_{k = 1}^K {\sqrt {\rho _{DL,lk}^D} {{\mathbf{w}}_{dk}}{\mathbf{x}}_{dk}^D} }$ where the first term is the residual inter-group data interference from cells other than ${{A}_{DI,l}}$ and the second term is the residual interference plus noise caused by the error of BS-BS channel estimation.

After the IC, the channel estimation of the $k'$-th MS in the $l$-th cell is given by
\begin{equation}
{\mathbf{\hat g}}_{llk'}^{IC} = {{\left( {{{{\mathbf{\bar y}}}_l}{\mathbf{\psi }}_{k'}^H} \right)} \mathord{\left/
 {\vphantom {{\left( {{{{\mathbf{\bar y}}}_l}{\mathbf{\psi }}_{k'}^H} \right)} {\left( {{F_c}{\tau _P}\sqrt {\rho _{UL,lk'}^P} } \right)}}} \right.
 \kern-\nulldelimiterspace} {\left( {{F_c}{\tau _P}\sqrt {\rho _{UL,lk'}^P} } \right)}}.
\end{equation}

Then, the MSCEE of the IC-TSP can be derived as (16), where the approximation is caused by omitting the correlation between the precoding vector ${{\mathbf{w}}_{dk}}$ and the BS-BS channel ${{\mathbf{G}}_{ld}}$ (similar to the derivation in Appendix A), $\varepsilon _{llk',pilot}^{{}}$ and $\varepsilon _{llk',noise}^{{}}$ are the same as those in (3), $\varepsilon _{llk',data,others}^{{}}$ is the impact of inter-group data interference from cells other than ${{A}_{DI,l}}$, $\varepsilon _{llk',data,residual}^{IC}$ is the impact of residual interference after IC (which is caused by the interference during the BS-BS estimation stage), $\varepsilon _{llk',noise,residual}^{IC}$ is the impact of residual noise after IC (which is caused by the additional noise during the BS-BS estimation stage). The deriving process of above-mentioned expectations are similar to that in Appendix A.

Compared to the TSP, the IC-TSP reduces the dominating component in MSCEE, i.e., the DL data interference from ${{A}_{DI,l}}$, from $\varepsilon _{llk',data,{{A}_{DI,l}}}^{{}}=\frac{P_{DL}^{D}}{{{F}_{c}}{{\tau }_{P}}\rho _{UL,lk'}^{P}}\sum\limits_{d\notin {{A}_{p}},d\in {{A}_{DI,l}}}^{{}}{{{\alpha }_{ld}}}$ to $\varepsilon _{llk',data,residual}^{IC}+\varepsilon _{llk',noise,residual}^{IC}$. $\sum\limits_{b\ne d,b\in {{B}_{d}}}^{{}}{{{\alpha }_{BS,lb}}}$ in $\varepsilon _{llk',data,residual}^{IC}$ is generally much smaller than ${{\alpha }_{ld}}$ in $\varepsilon _{llk',data,{{A}_{DI,l}}}^{{}}$. This is because the distance between the $l$-th BS to BSs in BS group ${{B}_{d}}$ is larger than the distance between the $l$-th BS to the the $d$-th BS due to the reuse of pilot matrix. Therefore, the IC-TSP can reduce the MSCEE effectively.

Note that the BS-BS channel estimation error ${{\mathbf{E}}_{ld}}$ could be reduced by using the linear minimum mean square error (LMMSE) channel estimation (for the detailed channel estimation process, please see Theorem 1 in [20]). Compared to the LS channel estimation, the LMMSE channel estimation further utilizes the BS-BS spatial correlation and the SNR information to suppress the interference and noise. Since the LMMSE method is an evolution of the LS method, it can be expected that the insights derived from the LS method still hold true for the LMMSE method.

Based on the analysis in Sec. III-B, we conclude that the SINR of IC-TSP scheme can be improved significantly due to the reduced MSCEE. Using the IC-TSP, the UL SINR is given by $SINR_{lk'}^{IC,UL}=SINR_{lk'}^{UL}\left| _{\varepsilon _{llk'}^{{}}\to \varepsilon _{llk'}^{IC}} \right.$, where $f\left( x \right)\left| _{x\to x'} \right.$ denotes the operation of replacing $x$ of $f\left( x \right)$ by $x'$. Other SINRs for DL can be obtained similarly. Similar to the TSP, the SINRs are almost the same for UL, PD and CL transmission.

Although IC-TSP can improve the SINR, additional radio resources are needed for the BS channel estimation. As a comparison, for TSP, the UL spectral efficiency is given by $\upsilon _{lk'}^{UL}={{\varpi }_{P}}{{\log }_{2}}\left( 1+SINR_{lk'}^{UL} \right)$, where ${{\varpi }_{P}}=\left( 1-\frac{{{\tau }_{P}}}{{{T}_{c}}} \right)$ is the effective resource ratio of TSP [1]. For IC-TSP, the effective resource ratio is given by ${{\varpi }_{P}}{{\varpi }_{T}}$, where ${{\varpi }_{T}}=1-{{{\tau }_{BS}}\left( {{L}_{D\_main}}+1 \right)}/{{{F}_{C}}{{T}_{BS\_C}}}\;$ and ${{\tau }_{BS}}\left( {{L}_{D\_main}}+1 \right)$ is the additional pilot overhead needed by the BS-BS channel estimation. Therefore, the UL spectral efficiency of IC-TSP is given by $\upsilon _{lk'}^{IC,UL}\!\!\!=\!{{\varpi }_{P}}{{\varpi }_{T}}{{\log }_{2}}\!\left( \!1\!+\!SINR_{lk'}^{IC,UL}\! \right)$. The spectral efficiency for PD and CL can be obtained similarly.

On one hand, the spectral efficiency can be improved by IC-TSP since it can reduce MSCEE significantly. On the other hand, the resource overhead needed for the BS-BS channel estimation would degrade the spectral efficiency. Therefore, the performance of IC-TSP depends on various system parameters. Since the spectral efficiencies at UL, PD and CL transmissions are almost the same, the following analysis will only be conducted for UL.

\subsection{Impact of ${T_{BS\_C}}$}

The spectral efficiency of IC-TSP depends on the length of coherence time of BS-BS channel ${{T}_{BS\_C}}$ since ${{\varpi }_{T}}$ is proportional to ${{T}_{BS\_C}}$. For small ${{T}_{BS\_C}}$, ${{\varpi }_{T}}$ is small and the overhead needed for the BS-BS channel estimation may be so large that there will be not enough resources left for data transmission. Thus, the spectral efficiency of IC-TSP may be inferior to that of TSP. A lower bound of ${{T}_{BS\_C}}$, $T_{BS\_C}^{min}$ can be found, below which the spectral efficiency of IC-TSP is less than that of TSP. $T_{BS\_C}^{min}$ can be derived by solving $\upsilon _{lk'}^{IC,UL}=\upsilon _{lk'}^{UL}$ (i.e., ${{\varpi }_{P}}{{\varpi }_{T}}{{\log }_{2}}\left( 1+SINR_{lk'}^{IC,UL} \right)={{\varpi }_{P}}{{\log }_{2}}\left( 1+SINR_{lk'}^{UL} \right)$), given by
\setcounter{equation}{16}
\begin{equation}
T_{BS\_C,lk'}^{min} = \frac{{{\tau _{BS}}\left( {{L_{D\_main}}{+}1} \right)}}{{{F_C}\left[ {1 - \frac{{{{\log }_2}\left( {1 + SINR_{lk'}^{UL}} \right)}}{{{{\log }_2}\left( {1 + SINR_{lk'}^{IC,UL}} \right)}}} \right]}}.
\end{equation}

In (17), both ${{\tau }_{BS}}$ and $\frac{{{\log }_{2}}\left( 1+SINR_{lk'}^{UL} \right)}{{{\log }_{2}}\left( 1+SINR_{lk'}^{IC,UL} \right)}$ monotonically increase with $M$, thus $T_{BS\_C,lk'}^{min}$ increases with $M$. Moreover, both the numerator and the denominator are increasing functions of ${{L}_{D\_main}}$. However, the increment of the numerator with ${{L}_{D\_main}}$ is larger than that of the denominator in (17) since the increasing speed of ${{\log }_{2}}\left( 1+SINR_{lk'}^{IC,UL} \right)$ with ${{L}_{D\_main}}$ is lower than 1. Therefore, $T_{BS\_C,lk'}^{min}$ also increases with ${{L}_{D\_main}}$. It should be noted that $T_{BS\_C,lk'}^{min}$ derived from (17) is different from one MS to another. To this end, the average of $T_{BS\_C,lk'}^{min}$ over large numbers of random MS realizations should be used to guide the system design. The impacts of $M$ and ${{L}_{D\_main}}$ are the same for the average of $T_{BS\_C,lk'}^{min}$.

\subsection{Impact of $M$}

${{\varpi }_{T}}$ is also closely related to the pilot length ${{\tau }_{BS}}$. For the considered LS BS-BS channel estimation, ${{\tau }_{BS}}=M$. Therefore, ${{\varpi }_{T}}$ is inversely proportional to the BS antenna number $M$, while the SINRs increase with $M$. Hence, there exists an optimal value for BS antenna number, $M_{opt}^{{}}$, maximizing the spectral efficiency of IC-TSP. When $M$ is larger than $M_{opt}^{{}}$, the spectral efficiency will decrease as $M$ increases. Moreover, when $M$ is sufficiently large, it is possible that spectral efficiencies of IC-TSP may be lower than those of TSP, due to the resource overhead introduced by the BS-BS channel estimation. Hence, there is a cross point $M$ of IC-TSP and TSP, $M_{cross}^{{}}$, beyond which the spectral efficiency of IC-TSP will be inferior to that of TSP.

\subsection{Impact of sectorization}

When directional antennas are deployed at the BSs and signals are received (and transmitted) at only part of the angular space of each BS antenna, cell sectorization can be used to reduce the inter-cell interference and improve the system capacity [40]-[42]. In this paper, we consider an ideal sectorization where each cell is divided into $\delta $ sectors and each sector is served by $\frac{M}{\delta }$ BS antennas. The signals in the direction of target sector over MS-BS channel obtain antenna directivity gain of $\delta $ (i.e., the signals will be multiplexed by $\sqrt{\delta }$), while signals in other directions will be restricted to zero [48]. Differently, the signals in the direction of target sector over BS-BS channel obtain antenna directivity gain of ${{\delta }^{2}}$ since both the transmitter and the receiver are equipped with directional antennas. Furthermore, compared to the unsectorized case, the number of interferers is reduced by $\delta $ times.

For the TSP with sectorization, to obtain the estimate of wireless channel ${{\mathbf{g}}_{llk'}}$ , we should firstly divide the received pilot signal by $\sqrt{\delta }$ and then conduct the channel estimation as that in (2), i.e, ${{\mathbf{\hat{g}}}_{llk'}}=\frac{\mathbf{y}_{l}^{\sec }\cdot \mathbf{\psi }_{k'}^{H}}{{{F}_{c}}{{\tau }_{P}}\sqrt{\delta }\sqrt{\rho _{UL,lk'}^{P}}}$, where $\mathbf{y}_{l}^{\sec }$ is the received pilot signal when sectorization is adopted. The MSCEE ${{\varepsilon }_{llk'}}=\frac{\delta }{M}\mathbb{E}\left\{ {{\left\| \mathbf{\hat{g}}_{llk'}^{{}}-\mathbf{g}_{llk'}^{{}} \right\|}^{2}} \right\}$ can be approximated by using the similar analysis in Sec. III-A. Compared to the unsectorized case, the power of intra-group pilot interference (which is the signal over MS-BS channel) from each interferer in the direction of target sector will remain the same since the loss of the effective BS antenna and the antenna directivity gain cancel out. Therefore, the reduction of interferers will lead to the reduction of intra-group pilot interference $\varepsilon _{llk',pilot}^{{}}$ by $\delta $ times. The power of inter-group data interference (which is the signal over BS-BS channel) from each interferer in the direction of target sector will increase $\delta $ times since the antenna directivity gain is $\delta $ times higher than the loss of the effective BS antenna (because both transmitter and receiver are equipped with directional antennas). The sectorization also reduces the number of inter-group data interferers by $\delta $ times. As a result, using the sectorization, the inter-group data interference $\varepsilon _{llk',data}^{{}}$ keeps the same with that in the unsectorized case. Recall that the MSCEE is dominated by the inter-group data interference. Therefore, thus the MSCEE of TSP experiences a marginal reduction after sectorization. Next, we turn to the impact of the sectorization on the SINR. As seen from Appendix B, the power of target signal and the correlated interference from each interferer are quadratic functions about the number of effective BS antennas while the power of uncorrelated interference is only linearly proportional to the number of effective BS antennas. Considering the loss of the number of effective BS antennas and the benefit derived from the antenna directivity gain and the interferer cancelling gain, we conclude that the sectorization will reduce the power of target signal and the power of uncorrelated interference by $\delta $ times while the power of correlated interference by ${{\delta }^{\text{2}}}$ times. As a result, compared to the unsectorized case, the SINR will increase marginally when $M$ is small (where the interference is dominated by the uncorrelated interference) and will increase significantly by $\delta $ times when $M$ is large (where the interference is dominated by the correlated interference).

For the IC-TSP, the intra-group pilot interference $\varepsilon _{llk',pilot}^{{}}$ can be reduced by $\delta $ times while its proportion in the MSCEE of IC-TSP is larger than that in the TSP. Furthermore, the residual interference caused by the BS-BS channel estimation error, i.e., $\varepsilon _{llk',data,residual}^{IC}$, can be reduced by $\delta $ times. This is because both the number of the interferers during the BS-BS channel estimation stage, i.e., $\left| {{B}_{d}} \right|$, and the number of interferers generating the dominant interference during the MS-BS channel estimation stage, i.e., $\left| {{A}_{DI,l}} \right|$, can be reduced by $\delta $ times. Therefore, compared to the TSP, the MSCEE of IC-TSP can be reduced more significantly by the sectorization due to the reduction of $\varepsilon _{llk',data,residual}^{IC}$ and $\varepsilon _{llk',pilot}^{{}}$. The analysis of SINR is similar to that in the TSP. Due to the decrease in the MSCEE, the SINR will be always improved by the sectorization whether $M$ is small or large. Furthermore, when $M$ is small, ${{\omega }_{T}}$ approaches 1 and the impact of reducing the overhead for BS-BS channel estimation on ${{\omega }_{T}}$ is marginal. However, when $M$ is large, ${{\omega }_{T}}$ is significantly affected by the pilot overhead and the sectorization will lead to remarkable increase in ${{\omega }_{T}}$. As a result, the spectral efficiency of the IC-TSP will be improved more significantly when $M$ is large. Besides, the sectorization is also beneficial in reducing the backhaul overhead in the IC-TSP with D-RAN structure since fewer BSs have to exchange their DL data and precoding vectors. Therefore, the sectorization is more useful for the IC-TSP.

\subsection{CS based BS-BS channel estimation}

As analyzed above, the overhead of BS-BS channel estimation has a significant impact on the spectral efficiency of IC-TSP, especially when $M$ is large. Therefore, it is important to reduce this overhead [43]. Since the BS-BS channel is Ricean and spatially correlated, it is expected that the BS-BS channel has a sparse representation in the spatial-frequency domain [44]. To realize a sparse representation of BS-BS channel in the spatial-frequency domain by fully exploiting channel correlations, the discrete Fourier transform (DFT) can be employed as the sparsifying-basis. Let ${{\overline{\mathbf{G}}}_{ld}}$ denote the sparse representation of the BS-BS channel ${{\mathbf{G}}_{ld}}$, given by
\begin{equation}
{{\overline{\mathbf{G}}}_{ld}}={{\mathbf{A}}^{H}}\mathbf{G}_{ld}^{H}\mathbf{A},
\end{equation}
where $\mathbf{A}\in {{\mathbb{C}}^{M\times M}}$ is the unitary DFT matrix which follows $\mathbf{A}{{\mathbf{A}}^{H}}={{\mathbf{A}}^{H}}\mathbf{A}={{\mathbf{I}}_{M}}$. Using the DFT matrix, the received pilot signal in the spatial-frequency domain is given by
\begin{equation}
\overline{\mathbf{Y}}_{ld}^{BS}=\overline{\mathbf{P}}\ \!{{\overline{\mathbf{G}}}_{ld}}+{{\overline{\mathbf{J}}}_{ld}},
\end{equation}
where $\overline{\mathbf{Y}}_{ld}^{BS}={{\left( \mathbf{Y}_{ld}^{BS} \right)}^{H}}\mathbf{A}$, $\overline{\mathbf{P}}=\sqrt{\rho _{{}}^{BS-P}}{{\mathbf{P}}^{H}}\mathbf{A}$, and ${{\overline{\mathbf{J}}}_{ld}}=\mathbf{J}_{ld}^{H}\mathbf{A}$. In the spatial-frequency domain, ${{\overline{\mathbf{G}}}_{ld}}$ is approximately ${{S}_{ld}}$-sparse, i.e., it can be represented up to a certain accuracy $F\left( 0<F\le 1 \right)$ using ${{S}_{ld}}$ non-zero coefficients [45]. The sparsity level ${{S}_{ld}}$ increases as $F$ increases. Based on the CS theory, when the pilot length (i.e.,  the column number of the pilot matrix) ${{\tau }_{BS}}\ge {{S}_{ld}}{{\log }_{2}}\frac{M}{{{S}_{ld}}}$ and the sensing matrix $\overline{\mathbf{P}}$ has restricted isometry property (RIP), the sparse signal ${{\overline{\mathbf{G}}}_{ld}}$ can be reconstructed from $\overline{\mathbf{Y}}_{ld}^{BS}$ [45]-[46]. Here, the pilot length  is chosen to be ${{\tau }_{BS}}=\max \left( {{S}_{ld}} \right){{\log }_{2}}\frac{2M}{\max \left( {{S}_{ld}} \right)}$ $(l,d\in \left\{ 1,2,...,L \right\}).$  To ensure the RIP of sensing matrix $\overline{\mathbf{P}}$, the pilot matrix ${\mathbf{P}}$ is chosen as the complex Gaussian matrix and shared among BSs in advance [46]. Therefore, the estimation of ${{\overline{\mathbf{G}}}_{ld}}$ can be derived by using orthogonal matching pursuit (OMP) algorithm [47] and should be inversely transformed to spatial domain to derive the BS-BS channel estimate $\mathbf{\hat{G}}_{ld}^{CS}$.

\begin{table*}[]
\centering
\caption{Parameter Settings}\label{I}
\begin{tabular}{|c|c|c|c|}
\hline
\textbf{Parameter}            & \textbf{Value}              & \textbf{Parameter}                                                   & \textbf{Value}                        \\ \hline
Default cell number  & $L=37$             & MS  transmitter power                                       & 23 dBm                       \\ \hline
Default group number & $\Gamma $=7        & Symbol number per coherence time                             & $T_c$=185 {[23]}            \\ \hline
MS number each cell  & $K$=20             & Default pilot sequence length                               & ${{\tau}_{P}}=4$             \\ \hline
Cell radius          & ${{r}_{c}} $=500 m & DL data length                                              & ${{T}_{d}}=96$               \\ \hline
Protection radius    & ${{r}_{d}} $=20 m  & UL data length                                              & ${{T}_{u}}=85$               \\ \hline
Carrier frequency    & 2 GHz              & Sub-carrier number per coherence frequency                   & $F_c$=5 {[20]}             \\ \hline
Bandwidth            & 10 MHz             & Noise power for all transmitting stages                     & -174 dBm/Hz                  \\ \hline
Rician factor        & ${{k}_{T}}$=10     & Default coherence time of BS-BS channel        & ${{T}_{BS\_C}}=500{{T}_{c}}$ \\ \hline
Decay exponent       & $\eta $=3.8        & Default number of main DL data interfering cells & ${{L}_{D\_main}}=18$  \\ \hline
Shadow fading factor   & ${{\sigma }_{sh}}=8$ dB [34] & Spatial correlation coefficient                             & $\kappa =0.8$ [25] \\ \hline
BS transmitter power & 46 dBm             & Certain accuracy $F$ to describe approximate sparsity                                         & $F=99\%$ \\ \hline
\end{tabular}

\end{table*}

\begin{table*}[]
\centering
\caption{Evaluation of average normalized MSCEE and MSCEE composition for TSP scheme}\label{II}
\begin{tabular}{|c|c|c|c|c|c|c|c|}
\hline
                                                                                                                                                                                                                                                                    & \begin{tabular}[c]{@{}c@{}}channel estimation\\  + precoding method\end{tabular} & $\Gamma =1$ & $\Gamma =3$ & $\Gamma =4$ & $\Gamma =7$ & $\Gamma =9$ & $\Gamma =12$ \\ \hline
\multirow{4}{*}{\begin{tabular}[c]{@{}c@{}}average normalized\\ MSCEE $\frac{\mathbb{E}\left\{ {{\varepsilon }_{llk'}} \right\}}{{{P}_{TC}}}$\\ (dB)\end{tabular}}                                                                                                  & LS+MF (analytical)                                                               & 2.66               & 7.41               & 7.65               & 7.71               & 7.77               & 7.87                \\ \cline{2-8}
                                                                                                                                                                                                                                                                    & LS+MF                                                                            & 2.89               & 7.61               & 7.75               & 7.89               & 7.90               & 7.93                \\ \cline{2-8}
                                                                                                                                                                                                                                                                    & LMMSE+MF                                                                         & 2.07               & 6.78               & 6.98               & 7.21               & 7.24               & 7.31                \\ \cline{2-8}
                                                                                                                                                                                                                                                                    & LS+ZF                                                                            & 2.82               & 7.63               & 7.76               & 7.89               & 7.89               & 7.92                \\ \hline
\multirow{4}{*}{\begin{tabular}[c]{@{}c@{}}Dominance of \\ ${{\varepsilon }_{llk',data}}$ in ${{\varepsilon }_{llk'}}$, \\ i.e., $\frac{\mathbb{E}\left\{ {{\varepsilon }_{llk',data}} \right\}}{\mathbb{E}\left\{ {{\varepsilon }_{llk'}} \right\}}$\end{tabular}} & LS+MF (analytical)                                                               & \textbackslash{}   & 88.55\%            & 91.93\%            & 92.97\%            & 95.93\%            & 96.92\%             \\ \cline{2-8}
                                                                                                                                                                                                                                                                    & LS+MF                                                                            & \textbackslash{}   & 91.09\%            & 92.68\%            & 93.34\%            & 96.32\%            & 97.31\%             \\ \cline{2-8}
                                                                                                                                                                                                                                                                    & LMMSE+MF                                                                         & \textbackslash{}   & 92.40\%            & 92.97\%            & 93.72\%            & 96.71\%            & 97.71\%             \\ \cline{2-8}
                                                                                                                                                                                                                                                                    & LS+ZF                                                                            & \textbackslash{}   & 90.71\%            & 92.68\%            & 93.44\%            & 96.42\%            & 97.41\%             \\ \hline
\end{tabular}

\end{table*}

Similar to the LS estimation of the BS-BS channel, the estimated BS-BS channel can be written as $\mathbf{\hat{G}}_{ld}^{CS}={{\mathbf{G}}_{ld}}+\mathbf{E}_{ld}^{CS}$, where $\mathbf{E}_{ld}^{CS}$ is the BS-BS channel estimation error. $\mathbf{\hat{G}}_{ld}^{CS}$ can be utilized to regenerate the inter-group interferences, which then be canceled in the IC-TSP scheme. Compared to the LS estimation of the BS-BS channel, the CS based method can reduce the pilot length ${{\tau }_{BS}}$ from $M$ to $\max \left( {{S}_{ld}} \right){{\log }_{2}}\frac{2M}{\max \left( {{S}_{ld}} \right)}$. As a result, the resource overhead for the BS-BS channel estimation can be reduced and ${{\varpi }_{T}}$ can be increased, especially when $M$ is large. However, to improve the accuracy of the BS-BS channel estimation, larger $F$ is needed. This will in return increase the pilot overhead since the pilot overhead increases as ${{S}_{ld}}$ increases. Hence, there exists a tradeoff between the channel estimation accuracy and the pilot overhead. To reduce the pilot overhead significantly, the CS based BS-BS channel estimation will sacrifice the BS-BS channel estimation accuracy. As a result, the CS based method is more suitable when $M$ is large where the reduction of pilot overhead is more important than the BS-BS channel estimation accuracy.

Compared to the orthogonal pilot matrix based approach, the CS based approach only needs the pilot matrix to be complex Gaussian to ensure the RIP of the sensing matrix. For the massive MIMO systems, the precoded DL data can be seen as approximately complex Gaussian since the precoding vector is derived from approximately complex Gaussian channel estimate. Furthermore, the precoded DL data is more close to complex Gaussian with the increase of $M$, which can be proved by using central limit theorem. Therefore, the precoded DL data can be used as pilots for estimating the BS-BS channel in the CS-based method. With the precoded DL data served as pilots in the CS based IC-TSP, the pilot overhead for BS-BS channel estimation can be reduced from ${{\tau }_{BS}}\left( {{L}_{D\_main}}+1 \right)$ to ${{\tau }_{BS}}{{L}_{D\_main}}$.

Moreover, the sectorization and the CS based BS-BS channel estimation can be combined together to reduce both the MSCEE and the overhead for the IC-TSP. The sectorization can improve the spectral efficiency of the CS based IC-TSP whether $M$ is small or large due to the reduction in MSCEE, the increase in SINR and the reduction in pilot overhead. When $M$ is large, the CS based IC-TSP with sectorization achieves the highest spectral efficiency among the considered TSP schemes in this paper. However, when $M$ is small, the spectral efficiency of the CS based IC-TSP with sectorization is lower than that of the LS based IC-TSP with sectorization. This is because the MSCEE of the CS based BS-BS channel estimation is higher than that of the LS based BS-BS channel estimation while the impact of reduction in the pilot overhead is marginal when $M$ is small.

\section{Performance Evaluation}
Simulations are carried out to evaluate the performance of the proposed IC-TSP and verify our analysis. System configurations are shown in Table I. The power parameters are chosen according to the LTE-A standard [33]. If there is no special declaration, the uniform power allocation, the LS channel estimation, the MF precoding and detection are adopted in simulations. Since the MSCEE, SINR and spectral efficiency are still random due to the impact of shadow fading and MSs' location, we generate 10000 random realizations of MS locations and shadow fading profiles to provide the average performance in the simulations.

Firstly, Table II shows the channel estimation performance of TSP with different channel estimation (LS and linear minimum mean square error (LMMSE)) and precoding methods (MF and ZF). The normalized MSCEE is defined as the ratio between the average MSCEE $\mathbb{E}\left\{ {{\varepsilon }_{llk'}} \right\}$ (average over all MSs) and the average power of the target channel ${{P}_{TC}}=\frac{1}{M}\mathbb{E}\left\{ {{\left\| \mathbf{g}_{llk'}^{{}} \right\|}^{2}} \right\}$ [27-28], [44]. At first, it can be seen that the analytical results of normalized MSCEE ( which are calculated using (3)) match well with the simulated ones, which verifies the validity of the approximated MSCEE. Considering the MSCEE performance with MF precoding and different channel estimation schemes, it can be seen that the average normalized MSCEE is larger than 6 dB for $\Gamma \ge 3$. This stands for an extremely high channel estimation error which deteriorates the system performance seriously. Meanwhile, the average normalized MSCEE increases rapidly when $\Gamma $ increases from 1 to 3, which verifies the analysis in Sec. III-A. Note that the gain in the average normalized MSCEE of LMMSE over LS is limited (smaller than 1dB). This is because the interference suffered by channel estimation of TSP is extremely severe. Next, the MSCEE performance with different precoding schemes is compared. It can be seen that with LS channel estimation, MF and ZF precoding present similar performance in MSCEE. This is because the precoding is designed to cancel the intra-cell interference (based on the MS-BS channel estimation) while the dominant component of MSCEE is the inter-group data interference (from the BS-BS channel), as also demonstrated in this table. Meanwhile, the composition of MSCEE is also evaluated. It can be seen that with LS/LMMSE channel estimation and MF/ZF precoding, the power of inter-group data interference $\mathbb{E}\left\{ {{\varepsilon }_{llk',data}} \right\}$ always dominates the MSCEE, which contributes more than 88\% to the total value. This verifies the analysis in Sec. III-A. In summary, for various TSP with LS and LMMSE channel estimations and MF and ZF precoding schemes, the system presents similar MSCEE performance, and the dominating MSCEE components are the same. Therefore, the analytical results obtained for TSP with LS channel estimation and MF precoding can also be insightful for TSP with LMMSE channel estimation and ZF precoding.

\begin{figure}[h]
  \centering
  \includegraphics[angle=0,width=0.5\textwidth]{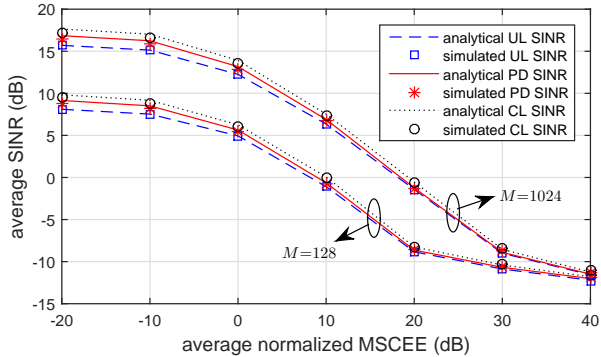}\\
  \caption{Average SINRs as a function of the average normalized MSCEE in TSP.}\label{Fig. 4}
\end{figure}

\begin{figure}[h]
  \centering
  \includegraphics[angle=0,width=0.5\textwidth]{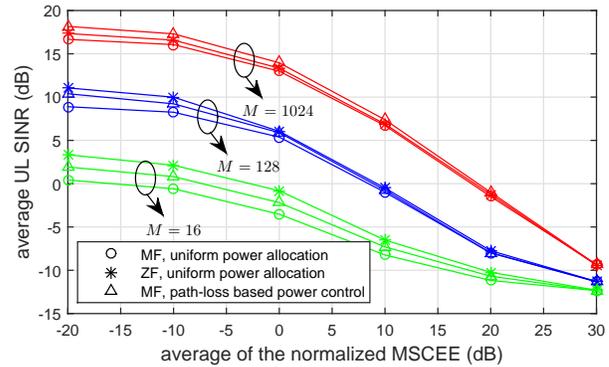}\\
  \caption{Average UL SINRs as a function of the average normalized MSCEE in TSP with different detection and power control.}\label{Fig. 5}
\end{figure}

Fig. 4 investigates the impact of average normalized MSCEE on the average SINRs of TSP where different MSCEE is derived by changing the DL data transmission power of interfering cells during the channel estimation of target cell. It can be seen that the analytical results (derived from (5), (7), and (8)) are quite close to the simulated ones. For the considered finite BS antenna cases, the UL SINR, PD SINR and CL SINR decrease with the increase of the average normalized MSCEE. The typical average normalized MSCEE from Table II with $\Gamma \ge 3$ is about 7.5 dB for LS channel estimation. When the average normalized MSCEE increases from -20 dB to this typical average normalized MSCEE, the UL SINR, PD SINR and CL SINR degrade by about 8 dB for both $M$=1024 and $M$=128. Therefore, it is important to improve channel estimation accuracy. Furthermore, the UL SINR, PD SINR and CL SINR are close to each other for different $M$ and MSCEE, which verifies the previous analysis in Sec. III. In the following simulations, the UL SINR is taken as an example to show the system performance.

The average UL SINR performance of TSP with ZF detection is shown in Fig. 5, and the impact of power control is also evaluated with MF method. It can be seen that the average UL SINR always decreases as the average normalized MSCEE increases, no matter which detection method and power control scheme are employed. When the average normalized MSCEE is small, using uniform power allocation, TSP with ZF method performs better than that with MF method, and the performance gain is larger with a smaller $M$. This is because the orthogonalization of MF method is strengthened with the increase of $M$ (which is called the asymptotic orthogonality in massive MIMO systems [1]) and the MF method approaches the performance of ZF method with larger $M$. Using the MF precoding, TSP with path-loss based power control performs better than that with uniform power allocation. The performance gap between path-loss based power control and uniform power allocation keeps stable as $M$ changes since the power allocation is independent with $M$. The previously mentioned performance gaps caused by different channel estimation and power control schemes reduce as the average normalized MSCEE increases. This is because TSP is trapped in severe channel estimation error with a large MSCEE and the advantages of ZF method and path-loss based power control become negligible. In summary, the insights derived from the analysis with MF method and uniform power control also hold for TSP with ZF method and path-loss based power control.

\begin{figure}[h]
  \centering
  \includegraphics[angle=0,width=0.5\textwidth]{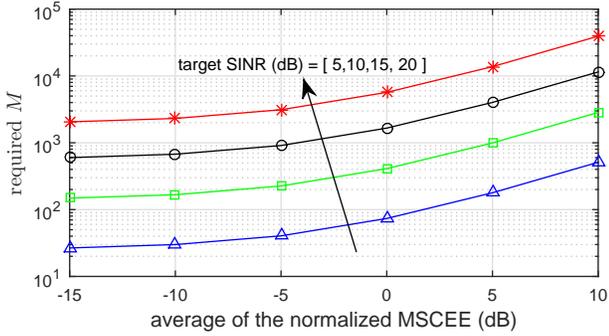}\\
  \caption{The required $M$ to achieve a target SINR as a function of average normalized MSCEE.}\label{Fig. 6}
\end{figure}

Fig. 6 shows the required $M$ (${{M}_{T}}$) to achieve a target SINR for UL transmission of TSP. ${{M}_{T}}$ increases rapidly with the average normalized MSCEE especially when the average normalized MSCEE is higher than 0 dB, which verifies the analyses in (10). To achieve a target SINR of 10 dB, TSP with the average normalized MSCEE of -10 dB requires about 170 antennas while the TSP with the average normalized MSCEE of 10 dB needs more than 2500 antennas, which becomes impractical for implementation. Hence, it is important to reduce MSCEE, so that less BS antennas are required to achieve the target performance.

\begin{figure}[h]
  \centering
  \includegraphics[angle=0,width=0.5\textwidth]{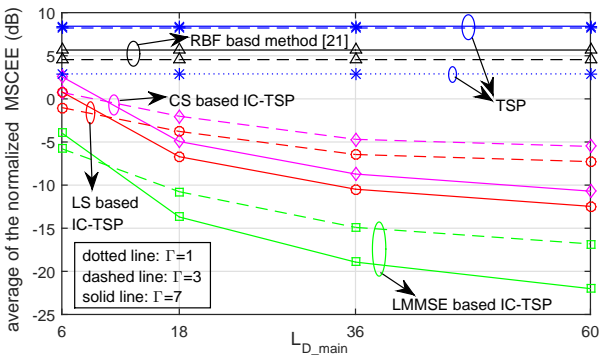}\\
  \caption{The average MSCEE as a function of ${{L}_{D\_main}}$.}\label{Fig. 7}
\end{figure}

Given $M$=128 and $L$=61, Fig. 7 shows the average normalized MSCEE performance of IC-TSP, TSP and RBF [21]. IC-TSP with different channel estimation schemes are evaluated, including LS (and LMMSE) based IC-TSP where all the BS-BS and MS-BS channels are estimated by LS (and LMMSE) method, and CS based IC-TSP where the BS-BS channels are estimated by CS based method and the MS-BS channels are estimated by LS method. IC-TSP always performs the best while the performance of traditional TSP is the worst. The RBF method in [21] outperforms TSP, thanks to the cancellation of UL data transmission. But its improvement is limited since the channel estimation error is dominated by DL data transmission but not UL data transmission. Furthermore, for IC-TSP, the average normalized MSCEE decreases as ${{L}_{D\_main}}$ increases since more inter-group data interference can be canceled. Considering the scenario with $\Gamma =7$, IC-TSP with LS channel estimation can reduce the average normalized MSCEE of the TSP by 15 dB and 19 dB for ${{L}_{D\_main}}=18$ and 36, respectively. The average normalized MSCEEs of TSP and the RBF method in [21] increase with $\Gamma $ since a higher $\Gamma $ leads to more DL data interference. However, the average normalized MSCEE of IC-TSP shows different trend with the change of $\Gamma $ for different ${{L}_{D\_main}}$. When ${{L}_{D\_main}}=6$, during the channel estimation, IC-TSP can only cancel the severe ICI generated from the nearest layer of cells transmitting DL data. So the channel estimation mainly suffers from the interference generated by the 12 cells in the second nearest layer. When $\Gamma =3$, 6 cells in this layer transmit DL data while another 6 cells transmit UL pilot. However, when $\Gamma =7$, all 12 cells in this layer transmit DL data. Since DL data interference is much higher than UL pilot interference, the interference with $\Gamma =3$ is less than that with $\Gamma =7$, and its channel estimation performance is better. Differently, when ${{L}_{D\_main}}=18$, IC-TSP with $\Gamma =7$ cancels interference from all 18 cells in the nearest two layers. However, IC-TSP with $\Gamma =3$ cancels interference from only 12 cells among the 18 cells, and the interference from the rest 6 cells transmitting UL pilot cannot be canceled. Therefore, the average normalized MSCEE of IC-TSP with $\Gamma =7$ become lower than that with $\Gamma =3$. When ${{L}_{D\_main}}$ increases from 18 to 36, the DL data interference generated from cells in the 3-rd layer is also cancelled. Due to the larger distance between the target cell and the cells in the 3-rd layer, the DL data interference is relatively small and the reduction in the MSCEE is limited. Furthermore, larger ${{L}_{D\_main}}$ leads to higher overhead. Thus, it is no need to apply ${{L}_{D\_main}}$ larger than 18. At last, for IC-TSP with different BS-BS channel estimation schemes, LMMSE based IC-TSP can achieve the lowest normalized MSCEE since the LMMSE BS-BS channel estimation utilizes the channel correlation information to reduce the interferences. Furthermore, the MSCEE of the LMMSE based IC-TSP shows the similar trend with that of the LS based IC-TSP. The CS based IC-TSP presents the highest average normalized MSCEE among these three IC-TSP schemes. As analyzed in Sec. IV-D, the BS-BS channel in the spatial-frequency domain is only approximately sparse (not strictly sparse), thus the BS-BS channel estimation error of CS based method is higher than LS and LMMSE method due to reconstruction error caused by the approximate sparsity.

\begin{figure}[h]
  \centering
  \includegraphics[angle=0,width=0.5\textwidth]{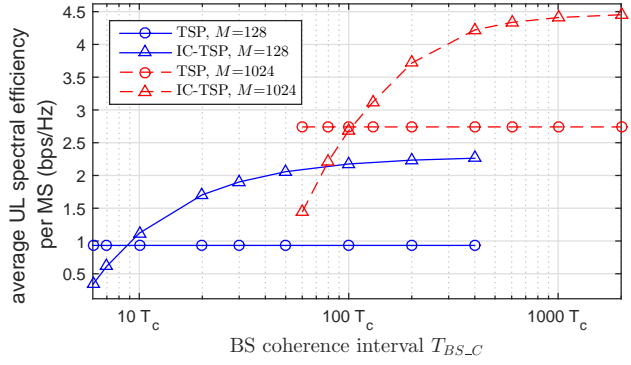}\\
  \caption{Impact of ${{T}_{BS\_C}}$ on average UL spectral efficiency.}\label{Fig. 8}
\end{figure}

Fig. 8 investigates the impact of the coherence time of the BS-BS channel ${{T}_{BS\_C}}$ on average UL spectral efficiency. It can be seen that the average UL spectral efficiency of the proposed IC-TSP increases with ${{T}_{BS\_C}}$ while the curves of TSP do not change with ${{T}_{BS\_C}}$. When ${{T}_{BS\_C}}$ is small, the overhead dominates the communication resources in ${{T}_{BS\_C}}$ and the overall performance is poor even if the channel estimation quality is improved. When ${{T}_{BS\_C}}$ increases, the impact of overhead decreases and the gains brought by good channel estimation become obvious. The cross point where IC-TSP exceeds TSP increases with $M$. This is because the overhead of BS-BS channel estimation increases with $M$. When ${{T}_{BS\_C}}$ increases further, the performance of IC-TSP keeps stable. This is because the overhead of BS-BS channel estimation becomes negligible for a large ${{T}_{BS\_C}}$, and the gain provided by improved channel estimation becomes saturated. When ${{T}_{BS\_C}}$ is large, it can be seen that IC-TSP achieves an average spectral efficiency gain of about 1.3 bps/Hz and 1.7 bps/Hz for $M$=128 and 1024, respectively.

\begin{figure}[h]
  \centering
  \includegraphics[angle=0,width=0.5\textwidth]{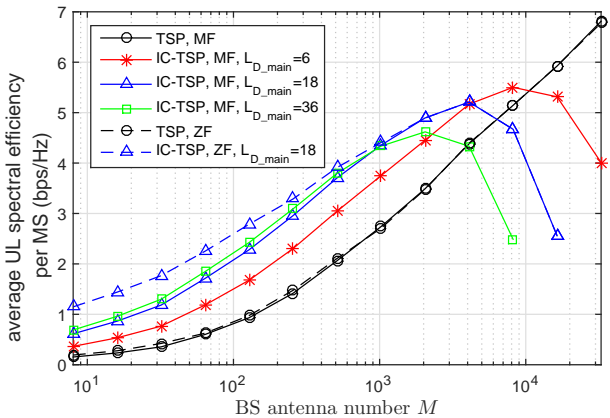}\\
  \caption{Impact of $M$ on average UL spectral efficiency.}\label{Fig. 9}
\end{figure}

\setcounter{TempEqCnt}{\value{equation}}
\setcounter{equation}{19}
\begin{figure*}[ht]
\begin{equation}
{\varepsilon _{llk'}} = \frac{{\mathbb{E}\left\{ {{{\left\| {\sum\limits_{j \ne l,j \in {A_p}}^{} {\sqrt {\frac{{\rho _{UL,jk'}^P}}{{\rho _{UL,lk'}^P}}} {{\mathbf{g}}_{ljk'}}} } \right\|}^2}} \right\} + \frac{{\mathbb{E}\left\{ {{{\left\| {\sum\limits_{d \notin {A_p}}^L {\sum\limits_{k = 1}^K {\sqrt {\frac{{\rho _{DL,dk}^D}}{{\rho _{UL,lk'}^P}}} {{\mathbf{G}}_{ld}}{{\mathbf{w}}_{dk}}{\mathbf{x}}_{dk}^D{\mathbf{\psi }}_{k'}^H} } } \right\|}^2}} \right\}}}{{{{\left( {{F_c}{\tau _P}} \right)}^2}}} + \frac{{\mathbb{E}\left\{ {{{\left\| {{\mathbf{n}}_l^P \cdot {\mathbf{\psi }}_{k'}^H} \right\|}^2}} \right\}}}{{\rho _{UL,lk'}^P{{\left( {{F_c}{\tau _P}} \right)}^2}}}}}{M},
\end{equation}

\hrulefill
\vspace{-0.5cm}
\end{figure*}

\setcounter{equation}{\value{TempEqCnt}}

Fig. 9 investigates the impact of $M$ on the average UL spectral efficiency considering both the MF and ZF method. At first, we concentrate on the MF method based TSP and IC-TSP scheme. It can be seen that when $M$ is small, IC-TSP always achieves higher average UL spectral efficiency than TSP. This is because IC-TSP achieves a much lower MSCEE and the overhead of BS-BS channel estimation is small when $M$ is small. The average spectral efficiency of TSP always improves with $M$ for the considered range. However, for IC-TSP, when $M$ is sufficiently large, it achieves the highest average spectral efficiency, then the average spectral efficiency decreases with $M$. This is because the overhead of BS-BS channel estimation increases linearly with $M$ and this reduces the spectral efficiency. A cross point occurs at certain $M$, beyond which the spectral efficiency of IC-TSP is worse than that of TSP. With ${{L}_{D\_main}}=18$, the optimal $M$ of IC-TSP is larger than 2048 and the cross point of $M$ is larger than 4096, which shows the effective range of $M$ for IC-TSP. The average spectral efficiency gain achieved by the IC-TSP with ${{L}_{D\_main}}=36$ is close to that with ${{L}_{D\_main}}=18$ while the optimal $M$ and cross-point $M$ is much smaller. Therefore, the spectral efficiency result also demonstrates that it is no need to apply ${{L}_{D\_main}}$ larger than 18. Comparing the IC-TSP with MF and ZF method, it is shown that IC-TSP with ZF method can achieve better performance than that with MF method when $M$ is smaller than 1000. When $M$ is large, the asymptotic orthogonality of massive MIMO system improves the performance of MF method and the gain of ZF method will vanish. However, for TSP, the spectral efficiencies of ZF and MF method are almost the same for all considered $M$. This is because the MSCEE is so large for TSP that the ZF method based on severely polluted channel estimation is hard to orthogonalize the multi-MS signals.

\begin{figure}[h]
  \centering
  \includegraphics[angle=0,width=0.5\textwidth]{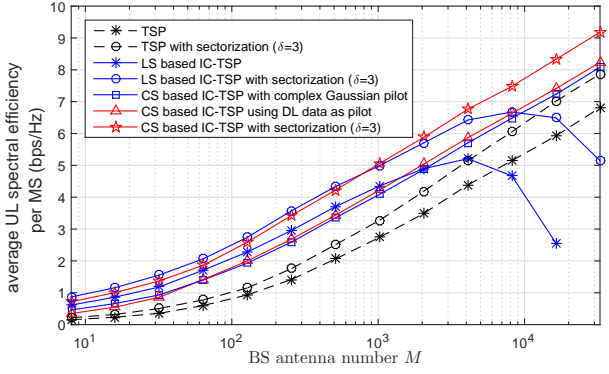}\\
  \caption{The average UL spectral efficiency as a function of $M$ , when TSP and IC-TSP are employed with various sectorization and channel estimation schemes.}\label{Fig. 10}
\end{figure}

Fig. 10 illustrates the UL spectral efficiency as a function of $M$, when TSP and IC-TSP are employed with various sectorization and channel estimation schemes. For IC-TSP, we set ${{L}_{D\_main}}=18$. Compared to the unsectorized case, the sectorization always improves the UL spectral efficiency while the improvement is more significant when $M$ is large. The difference is due to the fact that the correlated interference can be reduced more significantly than the uncorrelated one. Furthermore, compared to the TSP, the improvement of the UL spectral efficiency is more significant for the IC-TSP since the sectorization reduce the MSCEE more significantly in the IC-TSP. Next, we consider the performance of IC-TSP with CS based channel estimation (CS based IC-TSP). When $M$ is small, the average spectral efficiency of CS based IC-TSP is lower than that of LS based IC-TSP. This is because the ${{\varpi }_{T}}$ approaches 1 for both CS and LS based IC-TSP while the LS based IC-TSP achieves lower MSCEE. When M increases, the performance of CS based IC-TSP increases with $M$ even when $M=3\times {{10}^{4}}$. This is because using the CS based BS-BS channel estimation, the pilot overhead can be reduced significantly. As a result, the CS based IC-TSP achieves higher average spectral efficiency than LS based one for large $M$. In summary, it is recommended to use LS based IC-TSP for small $M$ (e.g., for $M\le 2000$ under the configuration in this paper) and CS based IC-TSP for large $M$(e.g., for $M>2000$ under the configuration in this paper). Furthermore, the spectral efficiency of the CS based IC-TSP using the precoded DL data as the pilot is also evaluated. The utilization of precoded DL data will slightly improve the spectral efficiency when $M$ is large due to the reduction of pilot overhead. However, the spectral efficiency will be reduced when $M$ is small where the RIP of sensing matrix cannot be well ensured. At last, we show the performance with the combination of the CS based BS-BS channel estimation and sectorization. When $M$ is large, this combination achieves the highest spectral efficiency. However, when $M$ is small, the spectral efficiency of the CS based IC-TSP with sectorization is lower than that of the LS based IC-TSP with sectorization, which is the same as that in the unsectorized case.

\section{Conclusions}
This paper focused on the finite antenna analysis for massive MIMO systems with TSP. After analytically demonstrating that the channel estimation error is critical for the system when the number of antenna is finite, an IC-based channel estimation method has been proposed in this paper. The main idea is to cancel out the inter-group data interference in the channel estimation, exploiting the shared information of precoding vectors, DL data and the estimated channels among BSs. The impacts of system parameters (including the length of the coherence time of BS-BS channel and $M$) and the pilot overhead reducing approaches (including the sectorization and the CS based BS-BS channel estimation) on IC-TSP have been extensively investigated. Both analytical results and simulations have shown that the proposed IC-TSP can effectively reduce the channel estimation error and improve the SINR and the spectral efficiency in the finite antenna massive MIMO system. For future work, the feasibility of machine learning based channel estimation should also be discussed for finite antenna massive MIMO systems to cancel the pilot contamination.


%

\appendices

\section{}

\setcounter{TempEqCnt}{\value{equation}}
\setcounter{equation}{20}
\begin{figure*}[ht]
\begin{equation}
{\varepsilon _{llk'}} \approx {\tilde \varepsilon _{llk'}} = \sum\limits_{j \ne l,j \in {A_p}}^{} {\frac{{\rho _{UL,jk'}^P}}{{\rho _{UL,lk'}^P}}{\beta _{ljk'}}}  + \frac{{P_{DL}^D}}{{{F_c}{\tau _P}\rho _{UL,lk'}^P}}\sum\limits_{d \notin {A_p}}^L {{\alpha _{ld}}}  + \frac{{\sigma _P^2}}{{{F_c} \cdot {\tau _P} \cdot \rho _{UL,lk'}^P}}.
\end{equation}
\end{figure*}

\begin{figure*}[ht]
\begin{equation}
SINR_{lk'}^{UL}\! =\! \frac{{\rho _{UL,lk'}^D\mathbb{E}\left\{ {{{\left\| {{\mathbf{\hat g}}_{llk'}^H{\mathbf{g}}_{llk'}^{}} \right\|}^2}} \right\}}}{{\left\{ \begin{gathered}
  \mathbb{E}\left\{ {{{\left\| {\sum\limits_{k = 1,k \ne k'}^K {\sqrt {\rho _{UL,lk}^D} {\mathbf{\hat g}}_{llk'}^H{\mathbf{g}}_{llk}^{}} } \right\|}^2}} \right\} + \mathbb{E}\left\{ {{{\left\| {\sum\limits_{j \in {A_p},j \ne l}^L {\sum\limits_{k = 1}^K {\sqrt {\rho _{UL,jk}^D} {\mathbf{\hat g}}_{llk'}^H{\mathbf{g}}_{ljk}^{}} } } \right\|}^2}} \right\} \hfill \\
   + \mathbb{E}\left\{ {{{\left\| {\sum\limits_{j \notin {A_p}}^L {\sum\limits_{k = 1}^K {\sqrt {\rho _{UL,jk}^D} {\mathbf{\hat g}}_{llk'}^H{\mathbf{g}}_{ljk}^{}} } } \right\|}^2}} \right\} + {{\mathbb{E}\left\{ {{{\left\| {{\mathbf{\hat g}}_{llk'}^H{\mathbf{n}}_{UL,lk'}^{}} \right\|}^2}} \right\}} \mathord{\left/
 {\vphantom {{\mathbb{E}\left\{ {{{\left\| {{\mathbf{\hat g}}_{llk'}^H{\mathbf{n}}_{UL,lk'}^{}} \right\|}^2}} \right\}} {{F_c}{T_u}}}} \right.
 \kern-\nulldelimiterspace} {{F_c}{T_u}}} \hfill \\
\end{gathered}  \right\}}}.
\end{equation}
\end{figure*}
\setcounter{equation}{\value{TempEqCnt}}

\setcounter{TempEqCnt}{\value{equation}}
\setcounter{equation}{23}
\begin{figure*}[ht]
\begin{equation}
\begin{array}{*{20}{l}}
  {\mathbb{E}\left\{ {{{\left\| {{\mathbf{e}}{{_{llk'}^{}}^H}{\mathbf{g}}_{llk'}^{}} \right\|}^2}} \right\}} \\
  { = \mathbb{E}\left\{ {{{\left\| {\sum\limits_{m = 1}^M {\left( {real\left( {e_{llk'm}^{}} \right) - j \cdot imag\left( {e_{llk'm}^{}} \right)} \right)\left( {real\left( {g_{llk'm}^{}} \right) + j \cdot imag\left( {g_{llk'm}^{}} \right)} \right)} } \right\|}^2}} \right\}} \\
  { = \sum\limits_{m = 1}^M \begin{gathered}
  \mathbb{E}\left\{ {rea{l^2}\left( {e_{llk'm}^{}} \right)} \right\}\mathbb{E}\left\{ {rea{l^2}\left( {g_{llk'm}^{}} \right)} \right\} + \mathbb{E}\left\{ {ima{g^2}\left( {e_{llk'm}^{}} \right)} \right\}\mathbb{E}\left\{ {ima{g^2}\left( {g_{llk'm}^{}} \right)} \right\} +  \hfill \\
  \mathbb{E}\left\{ {rea{l^2}\left( {e_{llk'm}^{}} \right)} \right\}\mathbb{E}\left\{ {ima{g^2}\left( {g_{llk'm}^{}} \right)} \right\} + \mathbb{E}\left\{ {ima{g^2}\left( {e_{llk'm}^{}} \right)} \right\}\mathbb{E}\left\{ {rea{l^2}\left( {g_{llk'm}^{}} \right)} \right\} \hfill \\
\end{gathered}  } \\
  { = M{\varepsilon _{llk'}}{\beta _{llk'}}}.
\end{array}
\end{equation}

\hrulefill
\vspace{-0.5cm}
\end{figure*}

\setcounter{equation}{\value{TempEqCnt}}


The MSCEE of the $k'$-th MS in the $l$-th cell is given by (20), where $\mathbb{E}\left\{ {{{\left\| {\sum\limits_{j \ne l,j \in {A_p}}^{} {\sqrt {\frac{{\rho _{UL,jk'}^P}}{{\rho _{UL,lk'}^P}}} {{\mathbf{g}}_{ljk'}}} } \right\|}^2}} \right\} = M\sum\limits_{j \ne l,j \in {A_p}}^{} {\frac{{\rho _{UL,jk'}^P}}{{\rho _{UL,lk'}^P}}{\beta _{ljk'}}}$ and $\mathbb{E}\left\{ {{\left\| \mathbf{n}_{l}^{P}\mathbf{\psi }_{k'}^{H} \right\|}^{2}} \right\}=M{{F}_{c}}{{\tau }_{P}}\sigma _{P}^{2}$.\\
However, the accurate result of $\mathbb{E}\left\{ {{\left\| \sum\limits_{d=1, d\notin {{A}_{p}}}^{L}{\sum\limits_{k=1}^{K}{\sqrt{\frac{\rho _{DL,dk}^{D}}{\rho _{UL,lk'}^{P}}}{{\mathbf{G}}_{ld}}{{\mathbf{w}}_{dk}}\mathbf{x}_{dk}^{D}\mathbf{\psi }_{k'}^{H}}} \right\|}^{2}} \right\}$ is hard to derive since the BS-BS channel ${{\mathbf{G}}_{ld}}$ is slightly correlated with the precoding vector ${{\mathbf{w}}_{dk}}$. Fortunately, this correlation is very weak. ${{\mathbf{w}}_{dk}}$ is generated by using the channel estimation ${{\mathbf{\hat{g}}}_{ddk}}$, where ${{\mathbf{\hat{g}}}_{ddk}}={{\mathbf{g}}_{ddk}}+{{\mathbf{e}}_{ddk}}$, ${{\mathbf{e}}_{ddk}}={{\mathbf{e}}_{ddk,pilot}}+{{\mathbf{e}}_{ddk,data}}+{{\mathbf{e}}_{ddk,noise}}$ and ${{\mathbf{e}}_{ddk,data}}=\frac{\left( \sum\limits_{b=1,b\notin {{A}_{p}}}^{L}\!\!{{{\mathbf{G}}_{db}}\sum\limits_{n=1}^{K}{\sqrt{\rho _{DL,bn}^{D}}{{\mathbf{w}}_{bn}}\mathbf{x}_{bn}^{D}}} \right)\cdot \mathbf{\psi }_{k}^{H}}{{{F}_{c}}{{\tau }_{P}}\sqrt{\rho _{UL,dk}^{P}}}$. In the numerator of ${{\mathbf{e}}_{ddk,data}}$, only the term with $b=l$ is correlated to ${{\mathbf{G}}_{ld}}$, which occupies $\frac{1}{L-\left| {{A}_{p}} \right|}$ of all the cumulated terms in the numerator of ${{\mathbf{e}}_{ddk,data}}$ where $L$ is the number of cells, ${{A}_{p}}$ is the set of pilot transmitting cells. Considering a common scenario described in Fig. 1 in this revision, $L=37$, $\left| {{A}_{p}} \right|=7$ thus the ratio $\frac{1}{L-\left| {{A}_{p}} \right|}$ is only $\frac{1}{30}$. Therefore, we omit this correlation and derive approximate analysis.  Furthermore, the BS-BS channel ${{\mathbf{G}}_{ld}}$ is given by ${{\mathbf{G}}_{ld}} = \sqrt {{\alpha _{ld}}} \left( {\frac{{\sqrt {{k_T}} }}{{\sqrt {{\text{1 + }}{k_T}} }}{{{\mathbf{\bar C}}}_{ld}}{\text{ + }}\frac{{\text{1}}}{{\sqrt {{\text{1 + }}{k_T}} }}{{\mathbf{C}}_{ld}}} \right)$, and $\mathbb{E}\left\{ {{\left\| \sum\limits_{d=1, d\notin {{A}_{p}}}^{L}{\sum\limits_{k=1}^{K}{\sqrt{\frac{\rho _{DL,dk}^{D}}{\rho _{UL,lk'}^{P}}}{{\mathbf{G}}_{ld}}{{\mathbf{w}}_{dk}}\mathbf{x}_{dk}^{D}\mathbf{\psi }_{k'}^{H}}} \right\|}^{2}} \right\}=$ $\frac{{{k}_{T}}}{1+{{k}_{T}}}\mathbb{E}\!\left\{ \! {{\left\| \sum\limits_{d=1, d\notin {{A}_{p}}}^{L}{\sum\limits_{k=1}^{K}\!{\sqrt{\frac{\rho _{DL,dk}^{D}}{\rho _{UL,lk'}^{P}}}\sqrt{{\alpha _{ld}}}{{{\mathbf{\bar{C}}}}_{ld}}{{\mathbf{w}}_{dk}}\mathbf{x}_{dk}^{D}\mathbf{\psi }_{k'}^{H}}} \right\|}^{2}}\! \right\}+$ $\frac{1}{1+{{k}_{T}}}\mathbb{E}\!\left\{\! {{\left\| \sum\limits_{d=1, d\notin {{A}_{p}}}^{L}{\sum\limits_{k=1}^{K}\!{\sqrt{\frac{\rho _{DL,dk}^{D}}{\rho _{UL,lk'}^{P}}}\sqrt{{\alpha _{ld}}}{{\mathbf{C}}_{ld}}{{\mathbf{w}}_{dk}}\mathbf{x}_{dk}^{D}\mathbf{\psi }_{k'}^{H}}} \right\|}^{2}}\! \right\}$, where the expectation in the first term equals to\!   $\mathbb{E}\!\left\{ \!{{\left\| \sum\limits_{d\notin {{A}_{p}}}^{L}{\sum\limits_{k=1}^{K}\!{\sqrt{\frac{\rho _{DL,dk}^{D}}{\rho _{UL,lk'}^{P}}}\sqrt{{{\alpha }_{ld}}}{{{\mathbf{\bar{C}}}}_{ld}}{{\mathbf{w}}_{dk}}\mathbf{x}_{dk}^{D}\mathbf{\psi }_{k'}^{H}}} \right\|}^{2}} \!\right\}\approx$ $ M{{F}_{c}}{{\tau }_{P}}\frac{P_{DL}^{D}}{\rho _{UL,lk'}^{P}}\sum\limits_{d\notin {{A}_{p}}}^{L}{{{\alpha }_{ld}}}$, and the expectation in the second term is $\mathbb{E}\left\{ {{\left\| \sum\limits_{d\notin {{A}_{p}}}^{L}{\sum\limits_{k=1}^{K}{\sqrt{\frac{\rho _{DL,dk}^{D}}{\rho _{UL,lk'}^{P}}}\sqrt{{{\alpha }_{ld}}}{{\mathbf{R}}^{\frac{1}{2}}}{{\mathbf{H}}_{W,ld}}{{\mathbf{R}}^{\frac{1}{2}}}{{\mathbf{w}}_{dk}}\mathbf{x}_{dk}^{D}\mathbf{\psi }_{k'}^{H}}} \right\|}^{2}} \right\}\approx$ $ M{{F}_{c}}{{\tau }_{P}}\frac{P_{DL}^{D}}{\rho _{UL,lk'}^{P}}\sum\limits_{d\notin {{A}_{p}}}^{L}{{{\alpha }_{ld}}}$. It can be seen that the impact of BS-BS interference is independent of the spatial correlation coefficient $\kappa$. This is because the spatial correlation does not impact the total power of interference. Thus, the MSCEE is approximated by (21).


\section{}

For the UL transmission stage, SINR of the detected signal of the $k'$-th MS in the $l$-th cell is given by (22). In (22), the channel estimation ${{\mathbf{\hat{g}}}_{llk'}}={{\mathbf{g}}_{llk'}}+\mathbf{e}_{llk'}^{{}}$. As shown in Fig. 1, when a target group transmits pilot in the $n$-th frame, the channel estimation is interfered by the precoded DL data of other groups. The precoding vectors of these interferences are generated using channel estimations conducted earlier, which are correlated with the channel estimation of the target group of the $\left( n-1 \right)$-th frame, but not those of the $n$-th frame. Since the wireless channels estimated at the $\left( n-1 \right)$-th and $n$-th frame are uncorrelated, ${{\mathbf{w}}_{dk}}$ in ${{\mathbf{e}}_{ll{k}',data}}$ is uncorrelated with ${{\mathbf{g}}_{ll{k}'}}$ corresponding to the $n$-th frame. As a result, in (2), ${{\mathbf{g}}_{ll{k}'}}$ and ${{\mathbf{e}}_{ll{k}'}}$ are uncorrelated. $\mathbb{E}\left\{ {{\left\| \mathbf{\hat{g}}_{llk'}^{H}\mathbf{g}_{llk'}^{{}} \right\|}^{2}} \right\}$ is given by
\setcounter{TempEqCnt}{\value{equation}}
\setcounter{equation}{25}
\begin{figure*}[ht]
\begin{equation}
SINR_{lk'}^{CL}{=}\frac{{\rho _{DL,lk'}^D\mathbb{E}\left\{ {{{\left\| {{\mathbf{g}}_{llk'}^T{{\mathbf{w}}_{lk'}}} \right\|}^2}} \right\}}}{{\left\{ {\begin{array}{*{20}{l}}
  {\sum\limits_{k = 1,k \ne k'}^K {\rho _{DL,lk}^D\mathbb{E}\left\{ {{{\left\| {{\mathbf{g}}_{llk'}^T{{\mathbf{w}}_{lk}}} \right\|}^2}} \right\}}  + \sum\limits_{j \ne l,j \in {A_p}}^L {\sum\limits_{k = 1}^K {\rho _{DL,jk}^D\mathbb{E}\left\{ {{{\left\| {{\mathbf{g}}_{jlk'}^T{{\mathbf{w}}_{jk}}} \right\|}^2}} \right\}} }  + } \\
  {\sum\limits_{j \in {A_q}}^{} {\sum\limits_{k = 1}^K {\rho _{UL,jk}^P\mathbb{E}\left\{ {{{\left\| {{g_{lk'jk}}} \right\|}^2}} \right\} + \sum\limits_{j \notin {A_p},j \notin {A_q}}^L {\sum\limits_{k = 1}^K {\rho _{DL,jk}^D\mathbb{E}\left\{ {{{\left\| {{\mathbf{g}}_{jlk'}^T{{\mathbf{w}}_{jk}}} \right\|}^2}} \right\}} }  + } } } \\
  { + {{\mathbb{E}\left\{ {{{\left\| {{\mathbf{n}}_{DL - CL,dk'}^{}} \right\|}^2}} \right\}} \mathord{\left/
 {\vphantom {{\mathbb{E}\left\{ {{{\left\| {{\mathbf{n}}_{DL - CL,dk'}^{}} \right\|}^2}} \right\}} {{F_c}{\tau _P}}}} \right.
 \kern-\nulldelimiterspace} {{F_c}{\tau _P}}}}
\end{array}} \right\}}},
\end{equation}
\end{figure*}
\setcounter{equation}{\value{TempEqCnt}}

\setcounter{TempEqCnt}{\value{equation}}
\setcounter{equation}{28}
\begin{figure*}[ht]
\begin{equation}
SINR_{lk'}^{CL} \approx \frac{{\left( {M + 1} \right)\beta _{llk'}^2 + \varepsilon _{llk'}^{}\beta _{llk'}^{}}}{{M\sum\limits_{j \ne l,j \in {A_p}}^{} {\frac{{\beta _{llk'}^{} + {\varepsilon _{llk'}}}}{{\beta _{jjk'}^{} + {\varepsilon _{jjk'}}}}\frac{{\rho _{DL,jk'}^D}}{{\rho _{DL,lk'}^D}}\frac{{\rho _{UL,lk'}^P}}{{\rho _{UL,jk'}^P}}\beta _{jlk'}^2} {+}\left( {\beta _{llk'}^{} + {\varepsilon _{llk'}}} \right){\varsigma _{CL,lk'}}}},
\end{equation}

\hrulefill
\vspace{-0.5cm}
\end{figure*}

\setcounter{equation}{\value{TempEqCnt}}

\setcounter{equation}{22}
\begin{equation}
\begin{gathered}
 \!\!\!\!\mathbb{E}\!\left\{ \!{{{\left\|{{\mathbf{\hat g}}_{llk'}^H{\mathbf{g}}_{llk'}^{}} \right\|}^2}}\! \right\}\!{=}\mathbb{E}\left\{\! {{{\left\| {{{\left( {{{\mathbf{g}}_{llk'}} + {\mathbf{e}}_{llk'}^{}} \right)}^H}{\mathbf{g}}_{llk'}^{}} \right\|}^2}} \right\} \hfill \\
   \quad\quad\quad\quad\quad\quad\quad\!\!\!\!\! {=} \mathbb{E}\!\left\{\! {{{\left\| {{\mathbf{g}}_{llk'}^H{\mathbf{g}}_{llk'}^{}} \right\|}^2}}\! \right\}\! +\! \mathbb{E}\!\left\{ {{{\left\| {{\mathbf{e}}_{llk'}^H{\mathbf{g}}_{llk'}^{}} \right\|}^2}}\! \right\}, \hfill \\
\end{gathered}
\end{equation}
where $\mathbb{E}\left\{ {{\left\| \mathbf{g}_{ll{k}'}^{H}\mathbf{g}_{ll{k}'}^{{}} \right\|}^{2}} \right\}=\mathbb{E}\left\{ \beta _{ll{k}'}^{2}{{\left\| \mathbf{h}_{ll{k}'}^{H}{{\mathbf{h}}_{ll{k}'}} \right\|}^{2}} \right\}=$ $\beta _{ll{k}'}^{2}\mathbb{E}\left\{ {{\left\| \mathbf{h}_{ll{k}'}^{H}{{\mathbf{h}}_{ll{k}'}} \right\|}^{2}} \right\}=\beta _{ll{k}'}^{2}\left( M+{{M}^{2}} \right)$ since $2\left\| \mathbf{h}_{ll{k}'}^{H}\mathbf{h}_{ll{k}'}^{{}} \right\|=\sum\limits_{m=1}^{M}{\left\{ {{\left[ \sqrt{2}\cdot real\left( h_{ll{k}'m}^{{}} \right) \right]}^{2}}+{{\left[ \sqrt{2}\cdot imag\left( h_{ll{k}'m}^{{}} \right) \right]}^{2}} \right\}}$ is a random variable follows Chi-squared distribution with $2M$ degrees of freedom, whose expectation is $2M$ and variance is $4M$. Furthermore, the channel estimation error ${{\mathbf{e}}_{llk'}}$ is uncorrelated with the target channel ${{\mathbf{g}}_{llk'}}$ (see (2)). Then $\mathbb{E}\left\{ {{\left\| \mathbf{e}_{llk'}^{H}\mathbf{g}_{llk'}^{{}} \right\|}^{2}} \right\}$ is given by (24). Therefore, $\mathbb{E}\left\{ {{\left\| \mathbf{\hat{g}}_{ll{k}'}^{H}\mathbf{g}_{ll{k}'}^{{}} \right\|}^{2}} \right\}=M\left( M+1 \right)\beta _{ll{k}'}^{2}+M\varepsilon _{ll{k}'}^{{}}\beta _{ll{k}'}^{{}}$. Similarly, other expectations in (22) can be derived as $\mathbb{E}\left\{ {{\left\| \sum\limits_{k=1,k\ne {k}'}^{K}{\sqrt{\rho _{UL,lk}^{D}}\mathbf{\hat{g}}_{ll{k}'}^{H}\mathbf{g}_{llk}^{{}}} \right\|}^{2}} \right\}=M\left( \beta _{ll{k}'}^{{}}+\varepsilon _{ll{k}'}^{{}} \right)\sum\limits_{k\ne {k}'}^{K}{\rho _{UL,lk}^{D}\beta _{llk}^{{}}}$, $\mathbb{E}\!\left\{\! {{\left\| \sum\limits_{j\in {{A}_{p}},j\ne l}^{L}{\sum\limits_{k=1}^{K}\!\!{\sqrt{\rho _{UL,jk}^{D}}\mathbf{\hat{g}}_{ll{k}'}^{H}\mathbf{g}_{ljk}^{{}}}} \right\|}^{2}}\! \right\}\!=$ $\!{{M}^{2}}\!\!\!\!\!\!\sum\limits_{j\in {{A}_{p}},j\ne l}^{{}}\!\!\!{\rho _{UL,j{k}'}^{D}\frac{\rho _{UL,jk'}^{P}}{\rho _{UL,lk'}^{P}}\beta _{lj{k}'}^{2}}+M\left( \beta _{ll{k}'}^{{}}\!+\!\varepsilon _{ll{k}'}^{{}} \right)\!\!\!\sum\limits_{j\in {{A}_{p}},j\ne l}^{{}}{\sum\limits_{k=1}^{K}{\rho _{UL,jk}^{D}\beta _{ljk}^{{}}}}$ (the correlation between the ${{\mathbf{e}}_{llk'}}$ and $\mathbf{g}_{ljk'}^{{}}$ leads to a correlated interference from the MSs using the same pilot sequence, whose power is $\mathbb{E}\left\{ {{\left\| \sum\limits_{j\in {{A}_{p}},j\ne l}^{L}{\sqrt{\rho _{UL,jk}^{D}}\mathbf{\hat{g}}_{ll{k}'}^{H}\mathbf{g}_{ljk'}^{{}}} \right\|}^{2}} \right\}={{M}^{2}}\sum\limits_{j\in {{A}_{p}},j\ne l}^{{}}{\rho _{UL,j{k}'}^{D}\frac{\rho _{UL,jk'}^{P}}{\rho _{UL,lk'}^{P}}\beta _{lj{k}'}^{2}}+M\left( \beta _{ll{k}'}^{{}}+\varepsilon _{ll{k}'}^{{}} \right)\sum\limits_{j\in {{A}_{p}},j\ne l}^{{}}{\rho _{UL,jk'}^{D}\beta _{ljk'}^{{}}}$; other MSs introduce uncorrelated interference with the power of $\mathbb{E}\left\{ {{\left\| \sum\limits_{j\in {{A}_{p}},j\ne l}^{L}{\sum\limits_{k=1,k\ne k'}^{K}{\sqrt{\rho _{UL,jk}^{D}}\mathbf{\hat{g}}_{ll{k}'}^{H}\mathbf{g}_{ljk}^{{}}}} \right\|}^{2}} \right\}\!=\! M\left( \beta _{ll{k}'}^{{}}\!+\!\varepsilon _{ll{k}'}^{{}} \right) \sum\limits_{j\in {{A}_{p}},j\ne l}^{{}}{\sum\limits_{k=1,k\ne k'}^{K}{\rho _{UL,jk}^{D}\beta _{ljk}^{{}}}}$), $\mathbb{E}\left\{ {{\left\| \sum\limits_{j\notin {{A}_{p}}}^{L}{\sum\limits_{k=1}^{K}{\sqrt{\rho _{UL,jk}^{D}}\mathbf{\hat{g}}_{ll{k}'}^{H}\mathbf{g}_{ljk}^{{}}}} \right\|}^{2}} \right\}\!=\!$ {$M\left( \beta _{ll{k}'}^{{}}\!+\!\varepsilon _{ll{k}'}^{{}} \right)\sum\limits_{j\notin {{A}_{p}}}^{L}{\sum\limits_{k=1}^{K}{\rho _{UL,jk}^{D}\beta _{ljk}^{{}}}}$} and $\mathbb{E}\!\left\{\! {{\left\| \mathbf{\hat{g}}_{ll{k}'}^{H}\mathbf{n}_{UL,l{k}'}^{{}} \right\|}^{2}}\! \right\}$ $= M{{F}_{c}}{{T}_{u}}\left( \beta _{ll{k}'}^{{}}+\varepsilon _{ll{k}'}^{{}} \right)\sigma _{UL}^{2}$. As a result, $SINR_{lk'}^{UL}$ is given by
\setcounter{equation}{24}
\begin{equation}
\begin{gathered}
  SINR_{lk'}^{UL} =  \hfill \\
  \frac{{\left( {M + 1} \right)\beta _{llk'}^2 + \varepsilon _{llk'}^{}\beta _{llk'}^{}}}{{M\!\!\!\!\!\sum\limits_{j \ne l,j \in {A_p}}^{}\!\! {\frac{{\rho _{UL,jk'}^D}}{{\rho _{UL,lk'}^D}}\frac{{\rho _{UL,jk'}^P}}{{\rho _{UL,lk'}^P}}\beta _{ljk'}^2}  + \left( {\beta _{llk'}^{} + \varepsilon _{llk'}^{}} \right){\varsigma _{UL,lk'}}}} \hfill \\
\end{gathered} ,
\end{equation}
where ${\varsigma _{UL,lk'}} = \sum\limits_{k \ne k'}^K {\frac{{\rho _{UL,lk}^D}}{{\rho _{UL,lk'}^D}}\beta _{llk}^{}}  + \sum\limits_{j = 1,j \ne l}^L {\sum\limits_{k = 1}^K {\frac{{\rho _{UL,jk}^D}}{{\rho _{UL,lk'}^D}}\beta _{ljk}^{}} }  + \frac{{\sigma _{UL}^2}}{{\rho _{UL,lk'}^D}} = \sum\limits_{j = 1}^L {\sum\limits_{k = 1}^K {\frac{{\rho _{UL,jk}^D}}{{\rho _{UL,lk'}^D}}\beta _{ljk}^{}} }  - \beta _{llk'}^{} + \frac{{\sigma _{UL}^2}}{{\rho _{UL,lk'}^D}}$.

\section{}
For the CL transmission stage, SINR of received signal at the $k'$-th MS in the $l$-th cell is given by (26), where $\mathbb{E}\left\{ {{\left\| \mathbf{g}_{ll{k}'}^{T}{{\mathbf{w}}_{l{k}'}} \right\|}^{2}} \right\}=\mathbb{E}\left\{ {{\left\| \frac{\mathbf{g}_{ll{k}'}^{T}\left( {{{\mathbf{\hat{g}}}}_{ll{k}'}}^{*} \right)}{\left\| {{{\mathbf{\hat{g}}}}_{ll{k}'}} \right\|}\  \right\|}^{2}} \right\}$ can be approximated by $\frac{\mathbb{E}\left\{ {{\left\| \mathbf{g}_{ll{k}'}^{T}{{{\mathbf{\hat{g}}}}_{ll{k}'}}^{*} \right\|}^{2}} \right\}}{\mathbb{E}\left\{ {{\left\| {{{\mathbf{\hat{g}}}}_{ll{k}'}} \right\|}^{2}} \right\}}$ when $M$ is large [9], whose tightness can be verified by numerical simulation when $M\!>\!100$. $\mathbb{E}\left\{ {{\left\| \mathbf{g}_{ll{k}'}^{T}{{{\mathbf{\hat{g}}}}_{ll{k}'}}^{*} \right\|}^{2}} \right\}=\mathbb{E}\left\{ {{\left\| \mathbf{\hat{g}}_{ll{k}'}^{H}\mathbf{g}_{ll{k}'}^{{}} \right\|}^{2}} \right\}=M\left( M+1 \right)\beta _{ll{k}'}^{2}+M\varepsilon _{ll{k}'}^{{}}\beta _{ll{k}'}^{{}}$. $\mathbb{E}\left\{ {{\left\| {{{\mathbf{\hat{g}}}}_{llk'}} \right\|}^{2}} \right\}$ is given by
\setcounter{equation}{26}
\begin{equation}
\begin{array}{*{20}{l}}
  {\mathbb{E}\left\{ {{{\left\| {{{{\mathbf{\hat g}}}_{llk'}}} \right\|}^2}} \right\}}&{{\text{ = }}\mathbb{E}\left\{ {{{\left\| {{{\mathbf{g}}_{llk'}}} \right\|}^2}} \right\} + \mathbb{E}\left\{ {{{\left\| {{{\mathbf{e}}_{llk'}}} \right\|}^2}} \right\}} \\
  {}&{ = M\left( {\beta _{llk'}^{} + \varepsilon _{llk'}^{}} \right).}
\end{array}
\end{equation}

Therefore, $\mathbb{E}\left\{ {{\left\| \mathbf{g}_{llk'}^{T}{{\mathbf{w}}_{lk'}} \right\|}^{2}} \right\}$ in (26) is given by
\begin{equation}
\mathbb{E}\left\{ {{{\left\| {{\mathbf{g}}_{llk'}^T{{\mathbf{w}}_{lk'}}} \right\|}^2}} \right\} \approx \frac{{\left( {M + 1} \right)\beta _{llk'}^2 + \varepsilon _{llk'}^{}\beta _{llk'}^{}}}{{\beta _{llk'}^{} + \varepsilon _{llk'}^{}}}.
\end{equation}

Similarly $\sum\limits_{k=1,k\ne {k}'}^{K}{\rho _{DL,lk}^{D}\mathbb{E}\left\{ {{\left\| \mathbf{g}_{ll{k}'}^{T}{{\mathbf{w}}_{lk}} \right\|}^{2}} \right\}}\approx \sum\limits_{k=1,k\ne {k}'}^{K}{\rho _{DL,lk}^{D}\beta _{ll{k}'}^{{}}}=\beta _{ll{k}'}^{{}}\left( \rho _{DL}^{D}-\rho _{DL,l{k}'}^{D} \right)$,\\
$\sum\limits_{j\ne l,j\in {{A}_{p}}}^{L}{\sum\limits_{k=1}^{K}{\rho _{DL,jk}^{D}\mathbb{E}\left\{ {{\left\| \mathbf{g}_{jl{k}'}^{T}{{\mathbf{w}}_{jk}} \right\|}^{2}} \right\}}}\approx M\sum\limits_{j\ne l,j\in {{A}_{p}}}^{{}}{\frac{\rho _{UL,lk'}^{P}}{\rho _{UL,jk'}^{P}}\frac{\rho _{DL,j{k}'}^{D}\beta _{jl{k}'}^{2}}{\beta _{jj{k}'}^{{}}+\varepsilon _{jj{k}'}^{{}}}}+\rho _{DL}^{D}\sum\limits_{j\in {{A}_{p}},j\ne l}^{{}}{\beta _{jl{k}'}^{{}}}$, \\$\sum\limits_{j\in {{A}_{q}}}^{{}}{\sum\limits_{k=1}^{K}{\rho _{UL,jk}^{P}\mathbb{E}\left\{ {{\left\| {{g}_{l{k}'jk}} \right\|}^{2}} \right\}}}=\sum\limits_{j\in {{A}_{q}}}^{{}}{\sum\limits_{k=1}^{K}{\rho _{UL,jk}^{P}\mu _{l{k}'jk}^{{}}}}$, $\sum\limits_{j\notin {{A}_{p}},j\notin {{A}_{q}}}^{L}{\sum\limits_{k=1}^{K}{\rho _{DL,jk}^{D}\mathbb{E}\left\{ {{\left\| \mathbf{g}_{jl{k}'}^{T}{{\mathbf{w}}_{jk}} \right\|}^{2}} \right\}}}\approx$ \\$ \sum\limits_{j\notin {{A}_{p}},j\notin {{A}_{q}}}^{L}{\beta _{jl{k}'}^{{}}\sum\limits_{k=1}^{K}{\rho _{DL,jk}^{D}}}=\rho _{DL}^{D}\sum\limits_{j\notin {{A}_{p}},j\notin {{A}_{q}}}^{L}{\beta _{jl{k}'}^{{}}}$ and $\mathbb{E}\left\{ {{\left\| \mathbf{n}_{DL-CL,dk'}^{{}} \right\|}^{2}} \right\}={{F}_{c}}{{\tau }_{P}}\sigma _{CL}^{2}$. Therefore, the SINR of CL stage is approximated by (29), where ${\varsigma _{CL,lk'}} = \frac{{\rho _{DL}^D}}{{\rho _{DL,lk'}^D}}\sum\limits_{j = 1,j \notin {A_q}}^L {\beta _{jlk'}^{}}  - \beta _{llk'}^{} + \sum\limits_{j \in {A_q}}^{} {\sum\limits_{k = 1}^K {\frac{{\rho _{UL,jk}^P}}{{\rho _{DL,lk'}^D}}\mu _{lk'jk}^{}} }  + \frac{{\sigma _{CL}^2}}{{\rho _{DL,lk'}^D}}$.

\ifCLASSOPTIONcaptionsoff
  \newpage
\fi



%

%
\begin{IEEEbiography}[{\includegraphics[scale=1.7, bb=15 11 100 65 ,clip,keepaspectratio]{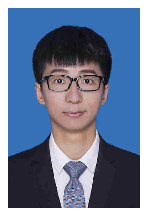}}]{Bule Sun}
received the B.S. degree in Communication Engineering from the Beijing University of Posts and Telecommunications in 2014. He is currently working toward the Ph.D. degree at the Institute of Computing Technology, Chinese Academy of Sciences. His current research interests include signal processing in massive MIMO systems, low resolution quantization and the convergence of communication and computing.
\end{IEEEbiography}

\begin{IEEEbiography}[{\includegraphics[width=1in,height=1.25in,clip,keepaspectratio]{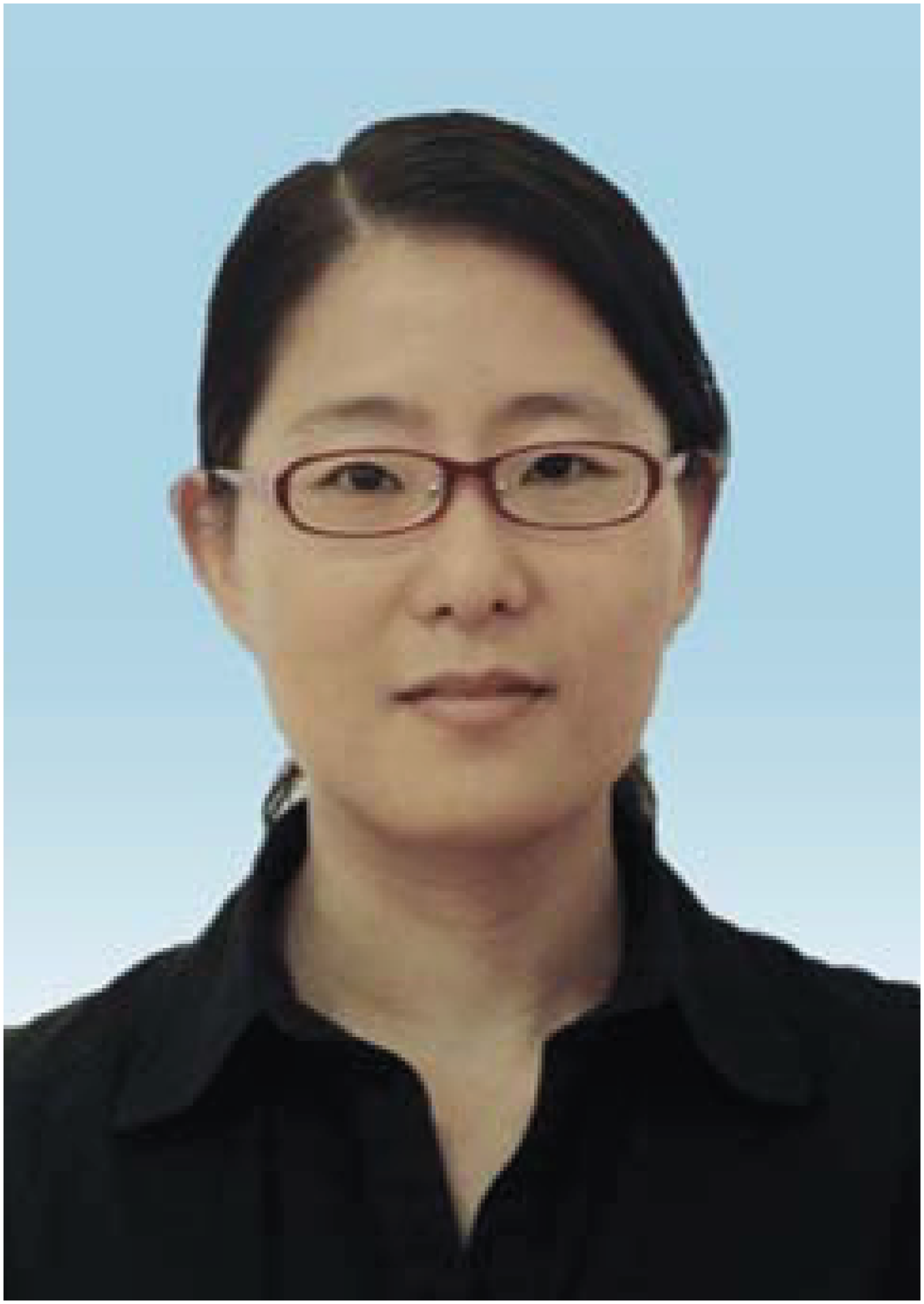}}]{Yiqing Zhou}
(S'03--M'05--SM'10) received the B.S. degree in communication and information engineering and the M.S. degree in signal and information processing from the Southeast University, China, in 1997 and 2000, respectively. In 2004, she received the Ph.D. degree in electrical and electronic engineering from the University of Hong Kong, Hong Kong. Now she is a professor in Wireless Communication Research Center, Institute of Computing Technology, Chinese Academy of Sciences. Dr. Zhou has published over 100 papers and four book/book chapters in the areas of wireless mobile communications. Dr. Zhou is the associate/guest editor for IEEE Trans. Vehicular Technology (TVT), IEEE JSAC (Special issue on Broadband Wireless Communication for High Speed Vehicles and Virtual MIMO), ETT and JCST. She is also the TPC co-chair of ChinaCom2012, symposia co-chair of IEEE ICC2015, symposium co-chair of GLOBECOM2016 and ICC2014, tutorial co-chair of ICCC2014 and WCNC2013, and the workshop co-chair of SmartGridComm2012 and GlobeCom2011. She received Best Paper Awards from ICC2018, ISCIT2016, IEEE PIMRC2015, ICCS2014 and WCNC2013. She also received the 2014 Top 15 Editor Award from IEEE TVT.
\end{IEEEbiography}

\begin{IEEEbiography}[{\includegraphics[width=1in,height=1.25in,clip,keepaspectratio]{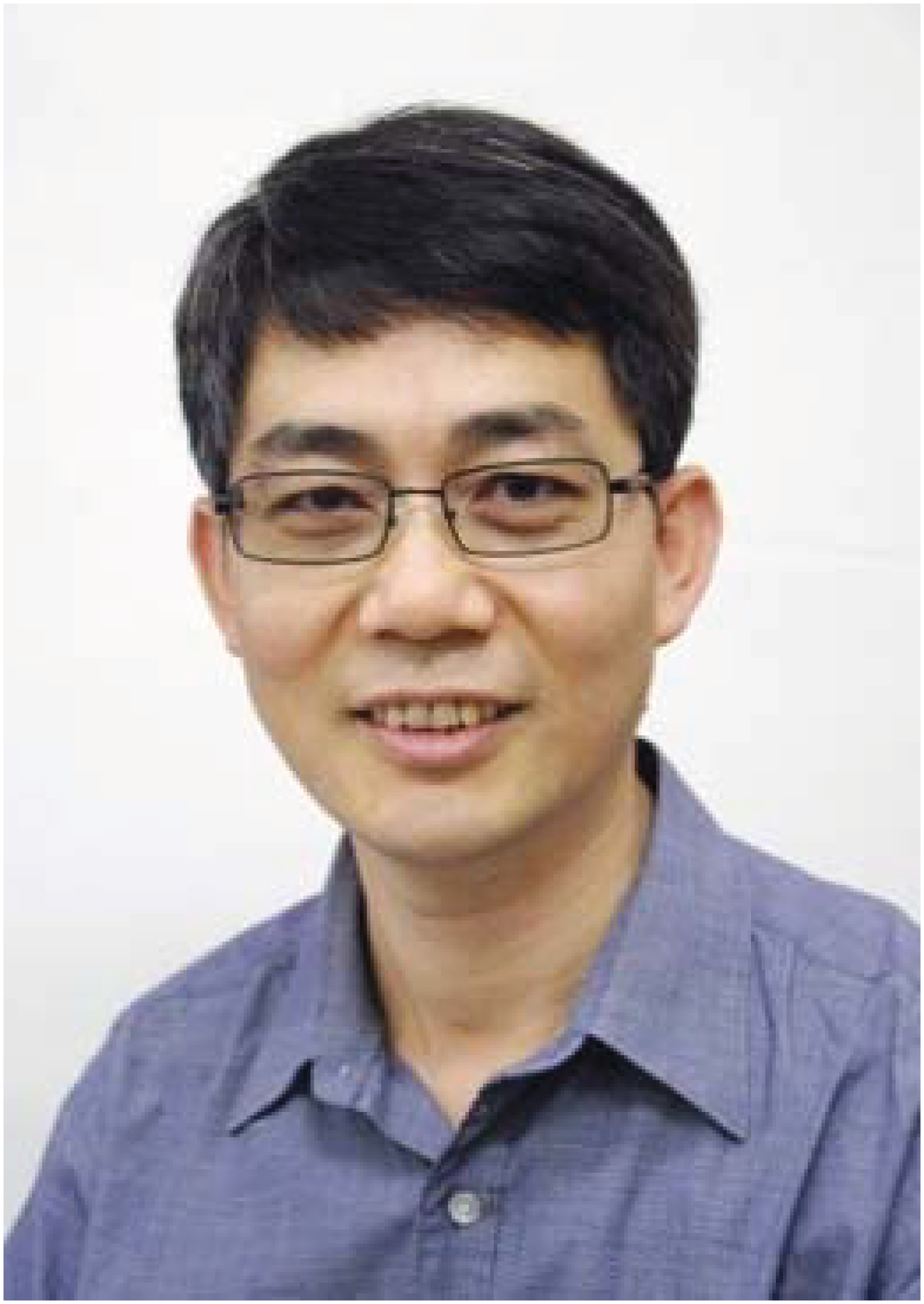}}]{Jinhong Yuan}
(M'02--SM'11--F'16) received the B.E. and Ph.D. degrees in electronics engineering from the Beijing Institute of Technology, Beijing, China, in 1991 and 1997, respectively. From 1997 to 1999, he was a Research Fellow with the School of Electrical Engineering, University of Sydney, Sydney, Australia. In 2000, he joined the School of Electrical Engineering and Telecommunications, University of New South Wales, Sydney, Australia, where he is currently a Professor and Head of Telecommunication Group with the School. He has published two books, five book chapters, over 300 papers in telecommunications journals and conference proceedings, and 50 industrial reports. He is a co-inventor of one patent on MIMO systems and two patents on lowdensity-parity-check codes. He has co-authored four Best Paper Awards and one Best Poster Award, including the Best Paper Award from the IEEE International Conference on Communications, Kansas City, USA, in 2018, the Best Paper Award from IEEE Wireless Communications and Networking Conference, Cancun, Mexico, in 2011, and the Best Paper Award from the IEEE International Symposium on Wireless Communications Systems, Trondheim, Norway, in 2007. He is an IEEE Fellow and currently serving as an Associate Editor for the IEEE Transactions on Wireless Communications. He served as the IEEE NSW Chapter Chair of Joint Communications/Signal Processions/Ocean Engineering Chapter during 2011--2014 and served as an Associate Editor for the IEEE Transactions on Communications during 2012--2017. His current research interests include error control coding and information theory, communication theory, and wireless communications.
\end{IEEEbiography}

\begin{IEEEbiography}[{\includegraphics[width=1in,height=1.25in,clip,keepaspectratio]{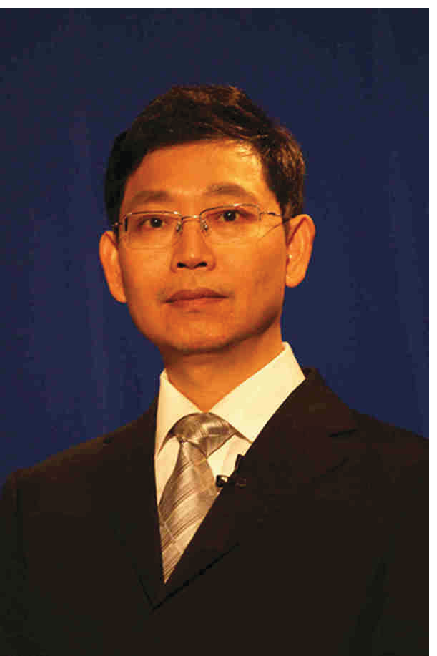}}]{Jinglin shi}
 is currently the director of the Wireless Communication Technology Research Center, ICT/CAS. He has published 2 books and more than 100 papers in telecommunications journals and conference proceedings, and has more than 30 patents granted. His research interests include wireless communication system architecture, signal processing, and baseband processor design. He was the General Co-Chair of ChinaCom'12, and a member of the TPC of IEEE WCNC, ICC, AusWireless2006, ISCIT 2007, and ChinaCom 2007 and 2009.
\end{IEEEbiography}








\end{document}